# Searching for memories, Sudoku, implicit check-bits, and the iterative use of not-always-correct rapid neural computation

**J. J. Hopfield**

Carl Icahn Laboratory, Princeton University, Princeton NJ 08544

**Abstract**

The algorithms that simple feedback neural circuits representing a brain area can rapidly carry out are often adequate to solve only easy problems, and for more difficult problems can return incorrect answers. A new excitatory-inhibitory circuit model of associative memory displays the common human problem of failing to rapidly find a memory when only a small clue is present. The memory model and a related computational network for solving Sudoku puzzles produce answers that contain implicit check-bits in the representation of information across neurons, allowing a rapid evaluation of whether the putative answer is correct or incorrect through a computation related to visual 'pop-out'. This fact may account for our strong psychological feeling of 'right' or 'wrong' when we retrieve a nominal memory from a minimal clue. This information allows more difficult computations or memory retrievals to be done in a serial fashion by using the fast but limited capabilities of a computational module multiple times. The mathematics of the excitatory-inhibitory circuits for associative memory and for Sudoku, both of which are understood in terms of 'energy' or Lyapunov functions, is described in detail.

**Prologue**
0n the other hand, a scientific paper could be written without a defined order to its intermediate sections. When the information framework is sufficiently complex, it is a vain folly to presume that there exists a unique order in which to best present the subject. No one reads a complex scientific paper from beginning to end, and in that single-pass processing has understood what the paper has to offer. Instead we read the abstract and part of the introduction, look at a few figures, read their captions, look for the essence of a particular experimental protocol, try to see the implications of an equation, glance at the conclusions, look at a figure again, … . We read a complex scientific paper in the same erratic and alogical way that we make sequential fixations with our eyes in understanding a complex visual scene. While I have sought a logical flow for this paper, the reader may prefer another ordering. The mathematically new results relate to a model of an associative memory based on an excitatory-inhibitory network, and the conceptually related mathematics of a Sudoku solver. They share the feature that the algorithm that the network carries out rapidly is not the ideal algorithm for the purpose, leading to errors (or illusions) in the answers provided by the network modules when the problems are difficult. Because erroneous answers contain implicit check-bit patterns that identify them as erroneous, they can be used as the basis for modifying the descending inputs to the module, so that the correct answer can be found by iterated use of the non-ideal algorithm. Feedback processes between brain modules permit the biological implementation of such iterative processing. Background lore sets these results in psychological, neurobiological, and computational context.

**Memory retrieval--personal**
Although I have thought long about the mathematics of associative memory, I look upon my own memory with a mixture of dismay and amazement. So often something comes up in a conversation that reminds me of a fact that I believe I know but just cannot get hold of at the moment. Who was the other leading actress in 'Some Like It Hot'? The conversation wanders to another topic, I converse with others, and half an hour later, while I am in the middle of a diatribe on the Bush administration, the name 'Shirley MacLaine' crashes into my conscious attention. It is as though a search had been commissioned, that search went on in a non-conscious fashion, and when it succeeded, the desired fact was then thrust into my consciousness and available for use. From the point of view of neural circuits and neural activity, what is taking place?

When such an item pops up into my conscious view it frequently 'does not feel quite right'. The name 'Shelly Winters' might have come to my conscious attention as a possible actress, which I would consciously reject as 'not feeling right'. When the non-conscious process finally throws up the suggestion 'Shirley MacLaine', I *know* 'that's it! No question. Right.' In neural terms, the signals developed in the non-conscious retrieval process can produce a strong and (reasonably) accurate feeling of correctness of the item retrieved. What is the basis for this calculation of confidence in correctness?

There are other seemingly related phenomena in the psychology of memory. When we cannot come up with a desired recollection immediately, we still seem to know whether we will soon be able to or not [Metcalf, 2000; Nelson, 1996]. Naming odors may also display a related phenomenon [Jonsson, 2005]. A typical response given when people are asked to name one particular familiar odor is "It is like—outdoors", rather than the actual name of the odor. However, told that the odor presented was grass, people then generally confirm this name with complete confidence.

Often when I try to recall a name, I first come up with a fragment, perhaps a feeling 'sounds like Taylor' without having the entire name yet available. If the name is required quickly, I may then even resort to going consciously through the alphabet Aaylor, Baylor, Caylor, Daylor ... . Suddenly, when I hit S, 'Sailor' *Aha*! I *know* that is the correct name. How can such surety of correctness come forth about something that I could not directly recall, rejecting other valid names like Taylor, Baylor, Mailer, ….? And how is this related to the previous paragraphs?

Is 'Silliam' a name? Our brains might contain a lookup table of names, and merely examine that list to answer the query. It seems more likely that most names are associated with individuals, and that we answer the question by seeing whether the clue 'Silliam' evokes the memory of some particular individual. The validity of the memory is important; the details of the invoked memory are not relevant.

**Is 'Aha is --correct!' a form of check-bit validation?**
Digital information can be corrupted in transmission, so check-bits or check digits are often included. Check-bits give the recipient the means to verify whether the information has arrived uncorrupted. As an example, in transmitting a 50 digit decimal number M, check-bits might consist of a 6 digit string N appended to the end of M, so that 56 digits are actually transmitted. These 6 digits N might be chosen as the middle 6 digits of $M^2$. N is then related to all the digits in M, but in so complex a fashion that N is essentially a random number based on M. A particular N will be generated by $10^{-6}$ of all possible examples of M. Thus if the number M N is received in data transmission and N agrees with what is calculated from $M^2$, the probability that there is *any* error in the transmission is no more than $10^{-6}$. Implementations of this idea with single or multiple check-bits or check-digits are in common use. The right-hand digit of most credit card numbers is a check-digit on the other numbers, calculated in such a way that any single digit mistake or transposition of adjacent digits can be recognized as an error.

The conceptual idea of check-bits for evaluating whether information has been corrupted has realizations in biology that are completely different from those of digital logic. How might a yeast cell check that a particular small protein molecule within the cell has been synthesized correctly from DNA instructions, i.e. that no substitution or frame-shift errors were made in the processes of DNA → mRNA → protein? There is no cellular mechanism for comparing in a 1:1 fashion the sequence of a protein molecule and all the coding sequences in the yeast genome to answer this question. Nor is the idea of check-digits a part of the information transfer processes in molecular biology. It might appear impossible that a cell could make such a correctness-check. For if proteins are thought of as random strings of amino acids, what basis could there possibly be to distinguish whether a random-looking sequence of amino acids has been synthesized in exact accord with some coding DNA sequence, or whether it has arisen as a result of an error?

However, proteins are *not* random sequences. Biological proteins fold into compact structures. Only compact folded proteins (which represent an infinitesimal fraction of all possible amino acid sequences) are biologically functional, so these are the sequences chosen through evolution. Most frame shift errors and a large fraction of substitution errors in synthesis result in proteins that are unable to fold so compactly. Checking that a protein has been synthesized without errors does not require knowing the correct sequence. A polypeptide that does not fold compactly is highly likely to be the result of an error in protein synthesis. Cells contain enzymes that recognize and destroy proteins that are not compactly folded. Thus, for biological proteins the ability of a protein molecule to fold compactly can be thought of as validating the check-bits for the protein, and molecules having incorrect check-bits are destroyed. In this case, unlike the digital example at the beginning of the section, the check-bits are implicit, and do not require the addition of extra amino acids at the end of a protein. Nor does understanding the correctness or incorrectness of the protein check-bits require Boolean logic. While I have somewhat overstated the effectiveness of this system for finding errors in proteins, I have done so only to point out how subtly the *idea* of check-bits can be implemented in real biology.

When the one brain area communicates the results of its 'computation' to another area, the signals could intrinsically contain a means by which the receiving area can test the incoming data for errors. This procedure need not necessarily utilize identified special transmission pathways or additional check-bits. It could be somehow intrinsic and implicit in the *pattern* of activity received, just as protein foldability is intrinsic to a sequence of amino acids. The receiving area could then identify the information as 'probably reliable' or 'probably erroneous', *even though the sending area might not have the ability to explicitly understand the reliability of the information it was sending.* If the decoding of the implicit (conceptual) check-bits pattern by the receiving area is rapid and produces a strong signal to higher-level processing, it would account for the strong "Aha!--correct" feeling we have about a difficult-to-recall item after it has been consciously or non-consciously sought and at long last correctly found.

**The modular brain**
Complex biological systems appear to be constructed from interacting modules. These modules are sometimes associated with physically separated groupings such as the neuron, or the replisome of molecular biology. A module has 'inputs' and 'outputs', and the details of how the inputs cause outputs within a module are chiefly hidden from the larger system. This allows the separate evolution of the capabilities of different modules, and also allows a description of the function of a higher level system in terms of modular function and models of modules. When a system is appropriately modular, new modules can often be added without the need to totally re-engineer the internal workings of existing modules. While interacting modules will certainly co-evolve to operate jointly, we can nevertheless *hope* to understand a biological system first in terms of stand-alone modules, and then to understand the dynamics of overall operation in terms of interacting modules exchanging signals [Hartwell, 1999].

The brain areas so evident to anatomy, electrophysiology, neurology, and psychology appear to be computational modules. These modules are intensely interconnected in an orderly fashion. Van Essen and coworkers [van Essen, 1985, 1992] have shown that there is a consistent and systematic definition of 'higher area' and 'lower area' in terms of which a visual area receives inputs from a lower cortical area into layer IV, while it receives inputs from higher cortical areas into layer I. The relationship of the visual cortex of simple mammals (e.g. the tree shrew with 8 visual areas [Lyon, 1998] and macaque monkeys ~ 32 neocortical visual areas [van Essen, 1992, 2004] can be understood on the basis of the evolutionary addition of more computing modules to higher animals. Because wiring rules are embedded in the molecular biology of nerve cells, there can be an automatic understanding of correct way to wire a new area into existing areas.

**Pre-attentive and attentive vision: pop-out**
The 'x' in the sea of 'gapped T' distractors, on the right-hand side Fig 1, 'pops out'. The 'x' is quickly perceived (in about 200 ms) in a time that is almost independent of the number of gapped T distractors.

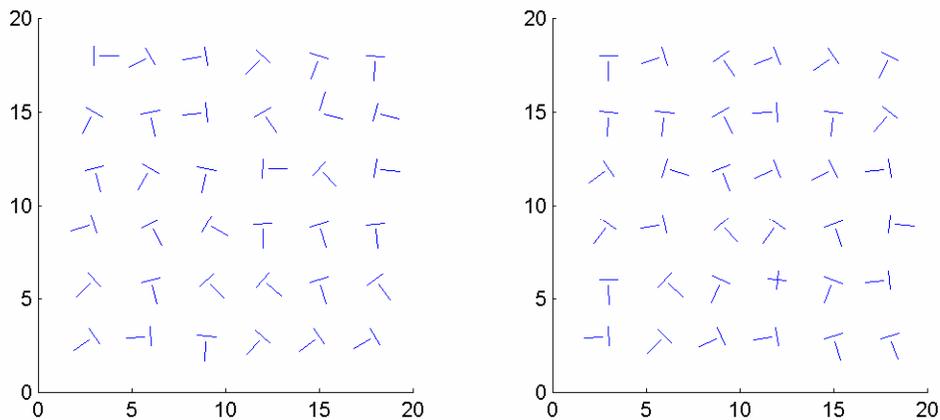

**Figure 1**. Single visual targets composed of two perpendicular equal-length line segments, amidst a sea of distractors composed from the same two perpendicular line segments. The distractors have the form of gapped T's (the letter T with a gap between the two line segments). The target on the right is a cross +, and pops out. The target on the left is a gapped L, and can be found only by sequential scrutiny.

By contrast, the time required to find a 'gapped L' embedded in a sea of 'gapped T' distractors (Fig 1,left side) has an approximately linear increase as the number of distractors is increased. The linear increase in the time required is attributed to the necessity of sequential scrutiny (or attention) of the items in the visual field [Treisman 1980, 1986]. Changing visual fixation can be a part of this sequential process, although the rapidity of the process indicates that several locations can be sequentially scrutinized within a single visual fixation.

The distinction between a discrimination which 'pops out' and one that does not is believed to be a distinction between a task that we do in parallel across a visual scene, and a task that requires sequential processing. When the parameters that draw the patterns to convert an 'x' to a 'gapped T' are changed in a continuous fashion, there is a narrow transition zone between pop-out and sequential attention. Such a transition zone is characteristic of psychophysical categories, whose idealization is a sharp boundary.

The phenomenon of pop-out is directly relevant to certain overall pattern classifications. Suppose that patterns for classification are chosen randomly from two classes. One class consists entirely of 'gapped T's. The other class of pattern consists chiefly of 'gapped T' and a few 'x' symbols. Because the 'x' symbols pop out, an immediate decision about the class of pattern will be possible in 200 ms. This kind of overall pattern discrimination is rapid, low level, does not require scrutiny, and is an elementary elaboration of the simple pop-out phenomenon.

**The fastest time scale of computation**
Some tasks of moderate complexity can be performed with great rapidity. A picture of a familiar face, presented in isolation on a blank background, can be recognized in 150 milliseconds. Rolls has argued (on the basis of the firing rates of neurons, the number of synapses and lengths of axons in the pathway between LGN and motor cortex, the known time delays between retina and LGN, and from motor cortex to action) that when this task is being performed as fast as possible, there is no time for feedback between visual areas, while there is time for feedback processes within each area [Panzeri 2001]. This rapid task can be done by cortical modules functioning in a feed-forward system, with multiple parallel paths contributing to the decision but without dynamical module-to-module feedback signals during the fastest recognition computations.

Within this fastest form of computation, without feedback between areas, visual brain areas can individually carry out useful algorithms which, when sequentially cascaded, can result in an appropriate action or answer. When a stimulus is presented, each module rapidly converges to a steady state determined by the stimulus. Once this convergence has finished, a module has ceased to perform computation. Computation in this short time-scale description is driven by the stimulus, and no additional computation will be done until the visual scene changes. The rapidness of pop-out perception indicates that it is due to a computation that can be carried out by a cascade of single convergences in early visual processing areas.

**Sequential processing for a fixed stimulus--chess**
Chess experts have an amazing feel for good moves.   An expert, looking at the board intuitively considers only a few possible moves, a few possible responses to those, etc. The size of the decision tree that the human expert explicitly explores in depth is small compared to that searched by an expert-system computer program.  A large part of the human expert's ability comes from the non-conscious, rapid, experience-based pattern recognition that prunes the exploration tree at each step.  Games like chess are interesting for neurobiology because they elicit, in a bounded yet difficult context, the kinds of processing necessary to do complex perceptual tasks.  The way humans play chess suggests that good chess players have acquired a non-conscious chess co-processor, and that this co-processor can be 'called' several times in a logical search for the best next move.

A chess-playing program organized in this fashion would have a very rapid pattern-recognition decision-tree pruning procedure or module acting at the level of one or two moves, and a higher level logical analysis. The higher level logical analysis would call the tree-pruning procedure several times in succession, positing 'if I move here, then my opponent has (from the pattern-recognition process) only 4 reasonable moves.  Examine each of them sequentially using the pattern-recognition procedure to suggest a few possibilities for my moves..... '.  The overall procedure would be based on sequentially exploring with logic a set of alternative choices generated by a heuristic pattern-recognition algorithm.

**What is sequential brain computation, and how does it come about?**
When there are sequential aspects of computation in the psychophysics of perception, some brain areas are presumably being used sequentially to do similar calculations. Attentional mechanisms in vision can refer to location, physical scale, eye of presentation, shape, … and to so many attributes that it is unlikely that the *choice* of attentional focus is made in or prior to V1.  It seems much more likely that higher processing areas create an attentional process that has a retrograde influence on earlier processing areas.  Such a process would have the effect of being able to use the simple processing and algorithmic capabilities of  a lower area many times in succession through sequential attentional focus, a powerful procedure will even when the sensory input is fixed.

While the scanning of the focus of visual attention might be controlled through input gating as in the 'searchlight' hypothesis [Crick 1982], there are much richer possibilities for using sequential top-down influences.  They may be providing additional informational signals to the lower area (for example, a position on a chess board due to a *hypothesized* move by an opponent.)  What the brain may be seeking could be a new computation by the lower area, using the same algorithm that it carries out when it is part of a feed-forward chain, but including the hypothesized data from a higher center.  The activity of lower-level visual areas when visualizing situations with closed eyes may exemplify such processing.  The response of neurons in V1 to illusory contours may be due to such signals.  Slow perceptual Gestalts such as the classic Dalmatian dog picture [James 1973] could easily involve several back-and-forths between higher areas suggesting 'try edges here and here' and a lower area seeing how things turn out with that suggestion plus the upward-flowing sensory information.

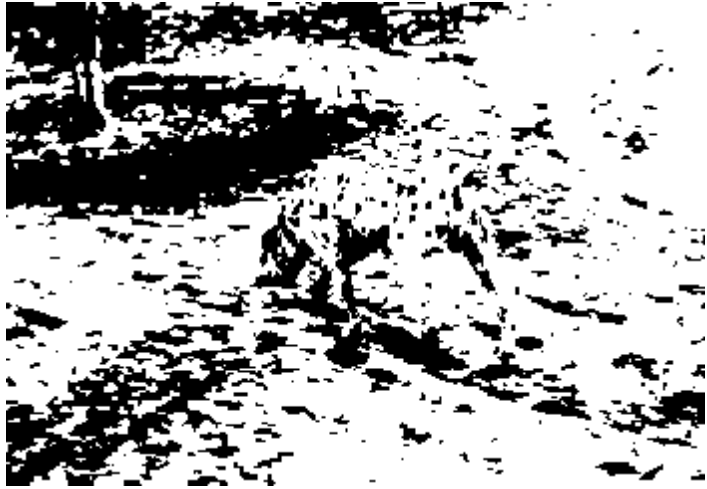

**Figure 2**. Find the Dalmatian. Hint: the dog is facing to the left, and the center of its chest is the dark area in the center of the picture.

### Effectiveness, algorithms, and illusions

For any special-purpose computational system to be effective in computation, there must be an excellent match between an algorithm employed and the hardware that carries out the computation. This is as true for neurobiological systems as for special-purpose digital computers. Illusions are symptoms of the fact that biology was not able to find an effective way to embed an always-correct algorithm in the neural system, but has instead used a less than perfect algorithm that more readily makes a hand-in-glove fit with neural hardware and information representations. The immense diversity of optical illusions indicates that the algorithms embedded in processing visual sensory information are riddled with assumptions, shortcuts, and approximations. Illusions are a product of the evolutionary tradeoff between effectiveness (speed, or amount of necessary hardware) and perceptual correctness.

Illusions are interesting because the errors represented by illusions can be more indicative of how processing actually takes place than correct processing would be. Calculating the digits of $\pi$ provides an analog to this situation. If program X generates answer $\pi = 3.1415926536..$ one learns little about the program from the answer. There are many ways to compute $\pi$ , and this correct answer does not tell us which was used. When program Y generates the answer $\pi = 3.1428571428571428571$ …, we can deduce that the algorithm was approximate, and that the approximate algorithm was $\pi = 22/7$.

### Universality classes and modeling neurobiology

Some behaviors of complicated systems are independent of details. The hydrodynamics of fluids results from strong interactions of the component molecules. Yet all fluids made up of small molecules have the same hydrodynamical behaviors; the details of the interactions disappear into a few macroscopic parameters such as the viscosity, compressibility, ... . The fluid behavior arising from small molecules is universal, while the notion of molecules has completely disappeared from the hydrodynamic equations.

Second order phase transitions occur in magnetism, fluids, mixtures, .. . The thermodynamical behavior of different second order transitions in the vicinity of the critical temperature falls into universality classes. Each class has a somewhat different behavior, but the behaviors within a class are the same. It is possible to understand which physical systems belong to which classes, and astonishingly different physical systems can belong to the same universality class and thus have the same mathematical structure near the phase transition.

Some dynamical systems also exhibit universal behavior close to special choices of parameters. The Hopf bifurcation is an example of such behavior, and the mathematics of the behavior of many systems goes to a universal form in the vicinity of such a bifurcation point.

I would never try to understand lift on an airplane wing using equations that described the motions and collisions of nitrogen molecules. Similarly, the modeling of this treatise is far simpler than real neurobiology. I believe that the only way we can get *systems understanding* in neurobiology is through such modeling. Unfortunately most notions of universality class are precise only in an infinitesimal region near a unique location in the state-space of a system. Moving way from such a point, each system begins to display its own unique features. My modeling of neurobiology as a dynamical system is done in the belief that there are semi-quantitative and phenomenological features that will be characteristic of a range of dynamical systems, so that in order to have an *understanding* it is necessary only to find a *simple model* that lies within that range. That model will not be in elementary correspondence with the details of electrophysiology, but the computational dynamics of the model will be a useful representation of the dynamics by which neurobiology carries out its computations.

**An associative memory within a biological setting**
*The memories*
We develop a modeling representation for a set of 'friends'. Each friend is described in terms the properties in a list of categories. The categories might include a first name, last name, age, city of birth, color of eyes, college attended, body build, city of residence, employer, political affiliation, name of partner, accent, favorite sport, .... .
.

| CATEGORY | property | property | property | property | property |
|---|---|---|---|---|---|
| FIRST NAME | john | mary | carlos | max | jessie |
| STATURE | very short | short | medium | tall | very tall |
| NATIONALITY | American | Mexican | English | Japanese | German |
| COLLEGE | Harvard | UCB | Caltech | Swarthmore | ETH |

For each of these categories there is a set of possible properties, as indicated in the grid above. A particular friend is then described by choosing entries in a grid of the same size.  Each entry (e. g. stature = medium) can be shared by many friends; it is the pattern of the entries that makes each friend unique.  Some friends have many properties; other friends are less well known, and have fewer 1's in their category/property grid.   Thus, for example, 4-year old Max, American, and not yet decided on college is described by the entries below.

| CATEGORY | property | property | property | property | property |
|---|---|---|---|---|---|
| FI⎮RST NAME | | | | 1 | |
| STATURE | 1 | | | | |
| NATIONALITY | 1 | | | | |
| COLLEGE | | | | | |

The mean activity level of neurons in association areas will depend upon the degree of specificity associated with a particular neuron.  If the activity of a particular neuron is assigned to a meaning as specific as a college or a first name, that neuron would seldom be active, and the representation of memories would have a sparseness beyond that expected from neurobiology.   It is much more likely that the representation is combinatorial. For example, several categories could be used to represent a name, with the first category referring to the first phoneme, and the property list for that category consist of all phonemes in the language.  The second category could be the second phoneme position, and have the same property list of all possible phonemes, and so on. A level of 10-100 properties for each category will provide some sparseness in representation, while retaining a reasonable mean level of neuronal activity when the neurons are activated in recall over the gamut of memories.

The actual properties of real friends often have probability distributions with long tails. For example, the fact that the populations of cities approximately follow Zipf's law (the population of the $n^{th}$ biggest city is $\sim 1/n$) means that the probability of being born in the $n^{th}$ largest city will scale as $1/n$.   Word usage probability also follows Zipf's law.  The occurrence of the 25 most popular names is approximately a power law, though with a form $1/n^{1/2}$ .

We choose for simulations a model that is mathematically simple, but incorporates a long-tailed power-law probability distributions.  Each memory has n_cat (usually taken as 50 in the simulations) different categories. Within each category, a memory has one of in_cat possible properties (usually taken as 20 in simulations).  It is convenient for the purposes of this paper to let the knowledge of each friend extend over all the categories. However, the model is capable of working with memories containing different amounts of information about the different friends, and has been shown to gracefully permit this flexibility with no qualitative alteration of its general properties.

For the modeling, a set of random friends is generated by randomly choosing in each category (for each friend) one property, e.g. one of the 20 possible values, according to a probability distribution p(n) $\sim 1/n^{1/2}$ .  The pattern of category/property entries for one typical friend is shown in Fig. 3, where the black rectangles represent the locations of 1's in a 50 by 20 table of categories and properties, a larger version of the table shown above.

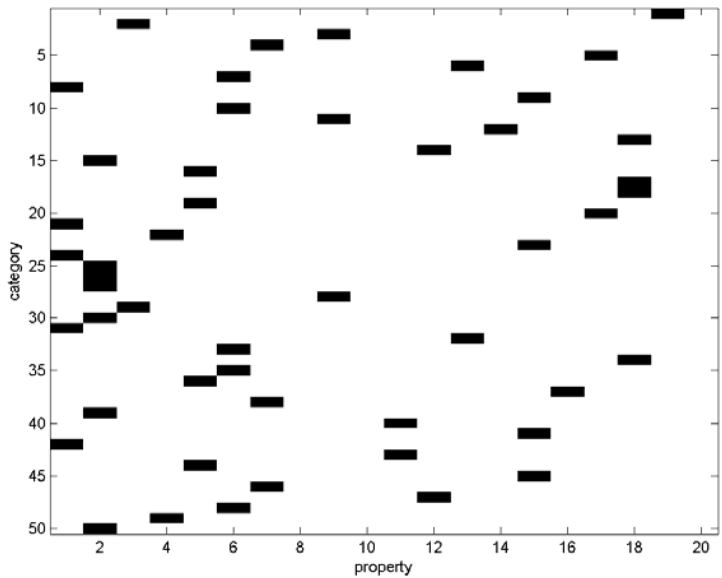

**Figure 3** Black rectangles indicate the activity pattern of 1000 neurons representing a typical memory. There are 50 categories, each with 20 possible values, and a memory has a single entry in each row representing a unique property in each category. The properties are listed in order of probability of occurrence. The visible bias of more 1's toward the left is due to the power law distribution over properties.

Neocortex has a sheet-like structure. Layer IV pyramidal cells make intralaminar excitatory synaptic contacts on a substantial fraction of their similar near neighbors, but are unlikely to make contacts with distant layer IV pyramidal cells. For short distance scales and small numbers of neurons, it makes sense to think of the system as allowing all-to-all synaptic connection possibilities, and an appropriate model has no geometry. At larger scales, the probability of connections drops off, the physical separation matters, and the cortex displays its sheet-like geometry and connection patterns. The modeling deals chiefly with a small patch of cortex having all-to-all possible connections. It is embedded in a larger sheet-like system that will be ignored except in the discussion. The maximum realistic size of such a patch might contain N = 1000-10000 pyramidal cells. In most of the simulations N = 1000 was used, with a single unit for each of the n_cat*in_cat possible entries in the property table. It makes no sense to take the large N limit of an all-to-all coupled system for present purposes. If it is at all permissible to first study a piece of intensely interconnected cortex and then consider its role in a larger system, the properties of the small system should be investigated for realistic values of N.

*The memory neural circuit*
The model neural circuit has all-to-all connections possible between N excitatory model neurons. In addition, an inhibitory network is equally driven by all N excitatory neurons, and in turn inhibits equally all excitatory units. A rate-based model is used, in which the instantaneous firing rate of each neuron is a function of its instantaneous input current. Each of the N locations in the property table is assigned to one neuron. Spiking neurons will be introduced only in discussion.

*The pattern of synapses*

The synaptic matrix T describing the connections between excitatory neurons for a set of memories is constructed as follows. Each synapse between two excitatory neurons has two possible binary values [Peterson 1998, O'Connor 2005], taken as 0 and 1. The two dimensional tableau representing the activity of a set of N = in_cat*n_cat neurons is an N-dimensional vector V. A new memory $V^{new}$ has components of 0 or 1 representing the property/value tableau of the new memory. This memory is inserted through changing the excitatory synapses T in a Hebbian fashion, by impressing the activity pattern $V^{new}$ on the system and then altering synaptic strengths according to the following rule.

*if* $V_j = 1$ and $V_k = 1$    make synapse $T_{jk} = 1$;  (regardless of its prior value of 0 or 1)
*else*                        make no change in $T_{jk}$.

Synapses of a neuron onto itself are not allowed. Starting with T = 0, the resulting symmetric synaptic matrix is independent of the order of memory insertion. T saturates at all 1's (except on the diagonal) when far too many nominal memories have been written, and then contains no information at all about the nominal memories. In the modeling, the number of memories $N_{mem}$ has been limited to a value such that the matrix T has about 40% 1's.

*The activity dynamics*

Excitatory neurons k have an input current $i_k$ resulting from exponential EPSC's given by

$$i_k(t) = \sum_p T_{kp} \int^t V_p \exp-(t'-t)/\tau_{ex} \, dt' + I_{in} \qquad (1)$$

$V_p$ is the instantaneous firing rate of excitatory cell p, $T_{kp}$ is the strength of an excitatory synapse from cell p to cell k, $\tau_{ex}$ is the effective excitatory time constant, and $I_{in}$ is the (negative) inhibitory current from the inhibitory network.

Since the inhibitory network receives equivalent inputs from all excitatory neurons, and sends equal currents to all excitatory neurons, from the modeling viewpoint it can be represented by a single 'neuron' whose input current from the excitatory EPSP's is

$$i_{toinhib} = \beta \sum_p \int^t V_p \exp-(t'-t)/\tau_{ex} \, dt' \qquad (2)$$

The firing rate V of an excitatory neuron is generically described by the function
V = g(i). In the simulations, V is given a specific form in terms of the input current i as

$$V(i) = \alpha^{ex}(i - i^{ex}_{thresh}) \quad \text{if} \ \ i > i^{ex}_{thresh} \qquad (3)$$
$$V(i) = 0 \qquad\qquad\qquad \text{if} \ i < i^{ex}_{thresh}$$

as sketched in the Fig. 4. The single inhibitory 'neuron' has its firing rate X given by

$$X = \alpha^{in}(i - i^{in}_{thresh}) \quad \text{if} \ \ i > i^{in}_{thresh} \qquad (4)$$
$$X = 0 \qquad\qquad\qquad \text{if} \ \ i < i^{in}_{thresh}$$

The following mathematics uses the semilinearity of the inhibitory network to achieve some of its simplicity. A more generic monotone increasing g(i) will suffice for the excitatory neurons.

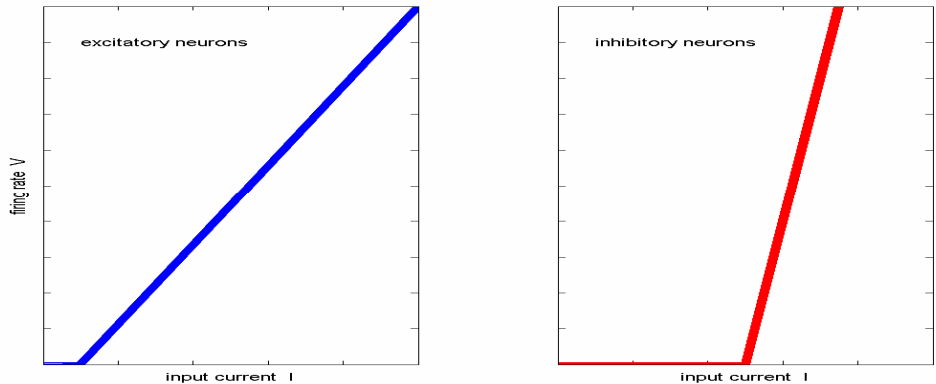

**Figure 4**. Firing rates vs. input current for excitatory and inhibitory model neurons.

Excitatory-inhibitory circuits of this general structure can oscillate. Oscillation can be avoided in a number of ways. We chose to prevent oscillation through using fast inhibition. In the limit of very fast inhibition

$$I_{in} = -\gamma\,(i - i^{in}_{thresh}) \quad \text{if} \quad i > i^{in}_{thresh} \;;\; 0 \text{ if } i \le i^{in}_{thresh} \tag{5}$$

$\gamma$ is a positive constant describing the strength of the inhibitory synapses. This scaling constant can be absorbed in the definition of $\alpha^{in}$, so can be set to 1 with no loss of generality. The function of the inhibitory network is to keep the excitatory system from running away, to limit the firing rate of the excitatory neurons. The only non-linearities in the network are the threshold non-linearities for both cell types.

In the fast inhibition limit, for a general g(i) but the specific form of h(i) the equations of motion of the activity of the system are

$$di_k/dt = -i_k/\tau_{ex} + \sum_j T_{kj}\,V_j - \alpha^{in}\beta\,h(\,\sum_j V_j - V_{tot})$$

$$h(V) \equiv V \text{ if } V{>}0 \text{ else } h(V) = 0 \text{ if } V{\le}0 \tag{6}$$

$$V_{tot} = i^{in}_{thresh}/\beta\tau_{ex}$$

The inhibitory term in the top line thus vanishes at low levels of excitatory activity.

*An energy function for the memory dynamics*

When T is symmetric, and the form of the inhibitory cell response is semilinear, a Lyapunov function E for the dynamics exists as long as the slope $\alpha$ of the inhibitory input-output relationship is large enough that the inhibition limits the total activity of the system to a finite value, and when g(V) is monotone increasing and bounded below. E defined below decreases under the dynamics. The derivation is closely related to the understandings of Lyapunov functions for simple associative memories [Hopfield 1982,1984] and for constraint circuits [Tank 1986].

$$E \;=\; 1/\tau_{ex}\sum_k \int^{Vk} g^{-1}(V')dV' \;-\; \tfrac{1}{2}\sum_{jk}T_{jk}\,V_jV_k \;+\; \beta\alpha^{in}/2\; H(\Sigma V_i - V_{tot})$$

$$H(V) \equiv V^2 \text{ if } V>0 \text{ else } H(V) = 0; \quad 1/2 \; dH/dV = h(V)$$

(7)

It is easiest to understand the stability implications of this function by first examining a limiting case, where $\tau_{ex}$ is very small, and $\beta\alpha^{in}$ is large. Then the function E being minimized by the dynamics is simply

$$E = -\tfrac{1}{2}\sum_{jk}T_{jk}V_jV_k \tag{8}$$

subject to the linear constraints

$$V_k \geq 0 \qquad \sum_j V_j < V_{tot} \tag{9}$$

The possible stable states of the system's N-dimensional space are values of V lying in a pyramid obtained by symmetrically truncating the (0,0,0, …) corner of an N dimensional hypercube by a hyperplane. Since T has a trace of zero, E cannot have its minimum value in the interior of this pyramid. All stable states must lie on the surface of this pyramid. However, they are generally *not* at the N+1 corners of the pyramid.

When the inhibition is not so fast, the system *may* oscillate. If the system does not oscillate, the stable states are independent of the time scale of the inhibition.

*Cliques and stable states*

Let a set of nodes be connected by a set of links. A subset of nodes such that each has a link to each other is called a clique. A clique is maximal if there is no node that can be added to make a larger clique. A clique can be described as a vector of 1's and 0's representing the nodes that are included. This allows a Hamming distance to be defined between cliques. A maximal clique will be called isolated if there is no other maximal clique at a Hamming distance 2 from it, i.e., when its neighboring cliques are all smaller.

The locations of the 1's in the connection matrix T define a set of links between the nodes (neurons). The stability properties of memories and non-memories can be stated in terms of cliques. It is easiest to understand the system when the slope of the inhibitory linear (above threshold) I-O relationship is very large. The value of the slope is in no way critical to the general properties, as long as it is sufficiently large to make the overall system stabilize with finite activity.   The simplest case is when $\tau_{ex}$ is very large and the first term in the Lyapunov function can be neglected.  However, finite time constant makes little significant difference unless the time constant is simply too small.

The following assertions can be easily illustrated and slightly more tediously proven

1) Any maximal isolated clique of T produces a stable state, *in which all the neurons in that clique are equally active.*
2) When two maximal cliques are adjacent (neither is isolated) the line between these two is also stable.
3) If T is generated by a single memory, the links define a single maximal isolated clique.  The precise number of 1's in the single memory does not matter.
4) When a few more desired memory states are added in generating T (as long as they are not extraordinarily correlated), T will have a structure that makes each of the desired memory states correspond to an isolated maximal clique.  This is true even when the different inserted memories have different numbers of non-zero entries.
5) As the number of states is increased, additional stable states of a different structure, not due to maximal cliques, arise.  These states are characterized by having differing values of neural activity across their participating neurons, unlike the case of an isolated maximal clique where all the activities are the same.
6) As more nominal memories are added, the system degrades when the number of memories is large enough that many nominal memories are no longer maximal cliques, or are maximal but not isolated and have several maximal neighbors.

The most significant features of this memory structure remain when $\tau_{ex}$ is finite, down to the point where if $\tau_{ex}$ is too small, the 'round the loop gain' can become less than 1, and the system loses its multistable properties.  In particular, for finite $\tau_{ex}$ and finite but sufficiently large $\alpha^{in}\beta$, isolated maximal cliques still define stable states in which all neurons in the clique are equally active.   When two maximal cliques are adjacent but $\tau_{ex}$ is finite, the previous line of stability between the two endpoints is replaced by a point attractor at the midpoint between these two cliques. (A simple ring attractor with a sequence of stable midpoints results from a diagonal strip of synapses in the present model.)

Thus the system *may* be a useful associative memory.  One cannot tell for sure until the basins of attraction are understood.  For that purpose, we turn to simulations.

*Simulations*
250 memories have been inserted in the example to be presented.   On average, each of the 1000 neurons is involved in 12.5 memories.  42% of the entries in T have been written as 1's.   A synapse for which  $T_{jh}$ =1 has on average been 'written' by 1.5 memories.  Of the 250 nominal memories, 239 are stable point attractors (isolated maximal cliques of T) of the dynamical system.   The other 11 nominal memory locations are from maximal cliques that are not isolated, and in each case (in the $\tau \rightarrow 0$ limit)  are at one end of a line-segment attractor.  The other end of the line segment attractor has a single '1' in the nominally stable memory *moved* to a location that should be a '0'.  At this level of loading, the memory scheme, with its simple write rule, is almost perfect in spite of strong correlations between memories.

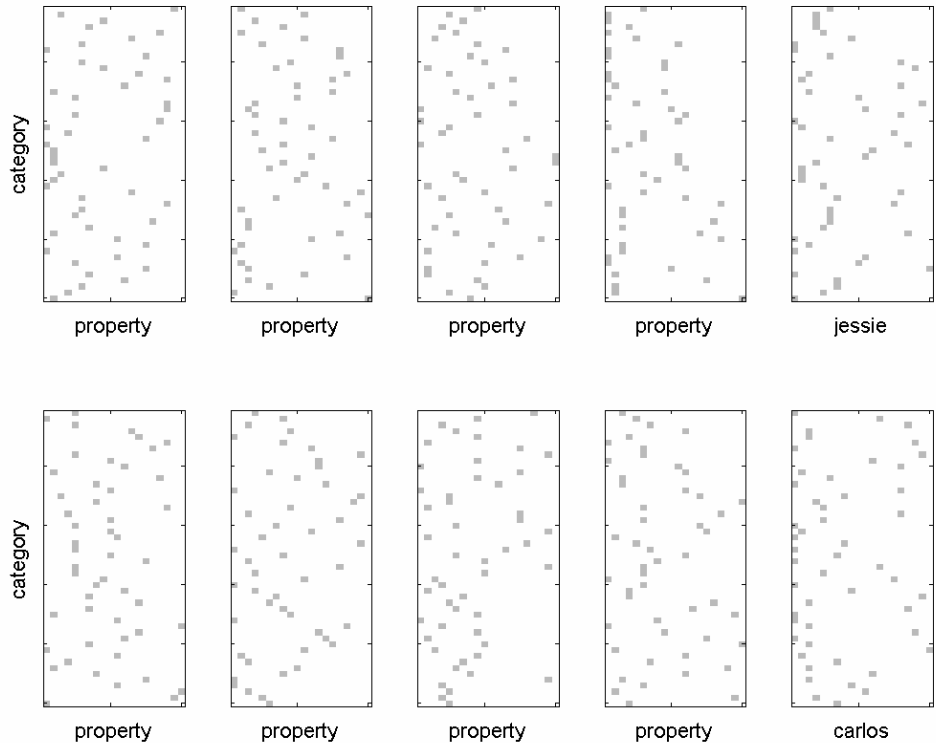

**Figure 5**. The first 10 of the 250 memories that are nominally embedded in the synaptic connections of the 1000 neurons system.  The representation is the same as that of Fig.3.

Fig. 5 shows the first 10 nominal memory states.  Starting with a pattern of u representing the nominal memory plus a random perturbation in each u, the system quickly returns to this point except for the case of the 11 memories that have line attractors.  In those cases, there is a maximal clique of the same size 2 units away in Hamming space, and the system settles at the midpoint of the states represented by the two adjacent maximal cliques.  In the subsequent figures, the memory load has been dropped to 225 for the purpose of simplifying the description, for at this slightly lower loading all the nominal memories are point attractors.

Fig. 6 shows the convergence of the system when 8 categories in the initial state are given the correct properties of an initial state, and there is no additional information.  This simulation used an initial state having $u_k = 1$ for 8 appropriately chosen values of k, and all other initial $u_k$ set to zero.  A detailed comparison of Fig. 6 with Fig 5 shows that the network is an accurate associative memory, reconstructing the missing 84% of the information with complete precision.

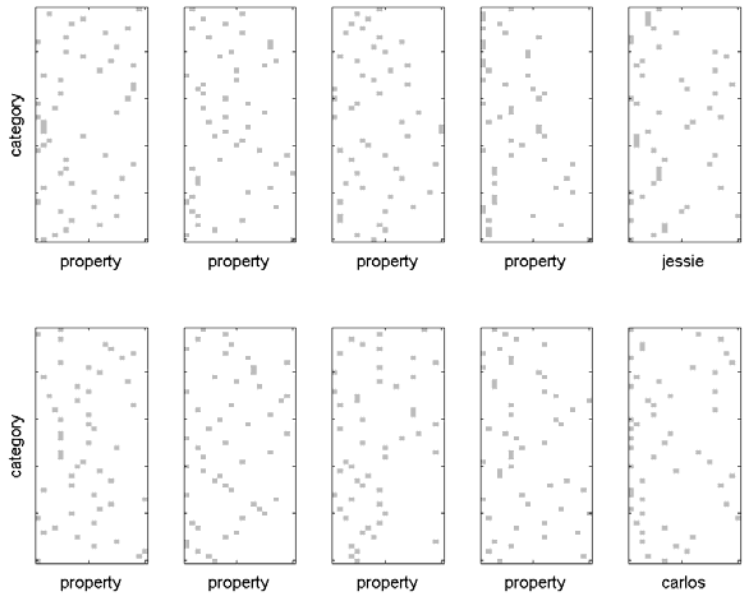

**Figure 6**. The first 10 of the 225 are tested for recall by starting in each case with an initial state representing correct property information about 8 categories, and 0's for all property entries in other categories. The results of the 10 separate convergences to final states are shown.

If, by contrast, we start with the components of u set randomly (u = 0.5*randn), the patterns found are very different, shown in Fig. 7. There are a large number of stable states of this system that are *not* memories. These junk states are qualitatively different from the memories from which T was made.

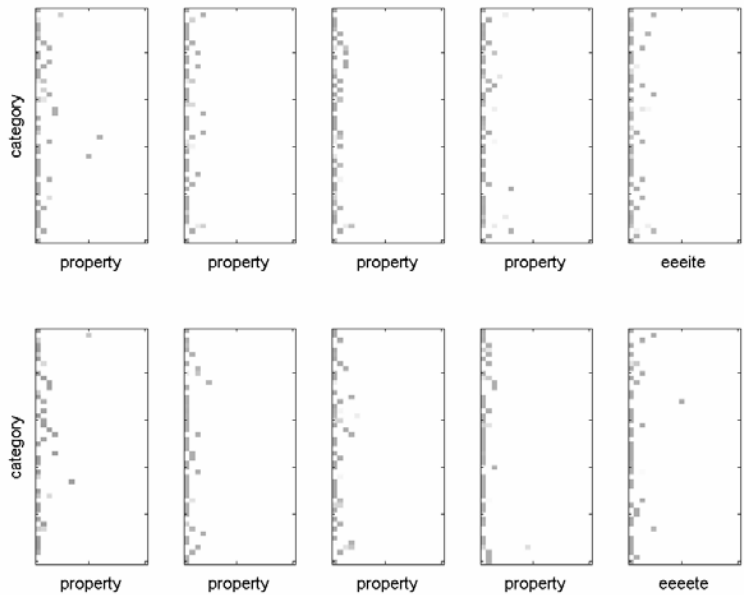

**Figure 7**. Results of 10 convergences from random starting states. None of the stable states that result are the nominal memory states or even qualitatively like the nominal memories.

*Memories of different lengths*

In the simulations, the memories have all been chosen to have the same total number of 1's, utilizing one property in each category. Nothing in the mathematics requires the cliques to have the same size, so memories having different lengths (memory length≡ total number of 1's) can be present in the same network. Simulations with lengths ranging over a factor of 2 confirm this. The level of activity of all the active neurons in a successful recall of a memory depends on the length of that memory. However, for each memory, the activities of all active neurons when recall is successful are equal. Since it is the equality of the levels of activity, not the level of activity itself, that identifies a correct memory, the correctness of a recollection can be visually verified even when a memory does not have a prior known length. When memories in a module have different lengths, the ease of recalling a memory correctly from a given query size depends on its length. In order not to deal with that complication in understanding simple results, the simulations presented here have all been done with memories of a common length.

*An 'I don't find anything' all-off state*

The effect of $i^{ex}_{thresh}$ can be understood in terms of equations of motion that pertain to the simple semilinear excitatory firing rate $g(i_k) = V_k = const*i_k$. Introducing a shift of $i^{ex}_{thresh}$ is equivalent to keeping the simple semilinear form and inserting instead a constant negative current into each excitatory neuron. Such a current adds a term $V i^{ex}_{thresh}$ to the Lyapunov function. The 'I don't find anything' state $\mathbf{V} = 0$ is then a stable state. The size of the attractor for this state depends on the size of $i^{ex}_{thresh}$, and this determines the amount of correct information necessary for a retrieval. In the present simulation, if $i^{ex}_{thresh} = 7$, the retrieval is exactly the same as in the example starting from a clue of 8 correct pieces of information above. However, when staring with the u set randomly (u = 0.5*randn), the dynamics take the system to the 'I don't find anything' state of zero activity. If the initial state is chosen with 7 pieces of information whose first-order statistics are the same as that for real memories, but which are not from one of the memories, the dynamics again take the system to the all-off stable state.

*Problems with retrieval from a weak correct clue*

In our personal use of our human associative memories, we generally begin from a clue that is chiefly or totally correct, about some small subset of the information in the memory. Our memories are harder to retrieve as the subset gets smaller. The following tableaux investigate this situation for the model, and show the results of retrieval from correct information in 8 categories (Fig. 6), in 6 categories (Fig. 8) and in 4 categories (Fig. 9) in the 50-category memory with $i^{ex}_{thresh} = 0$.

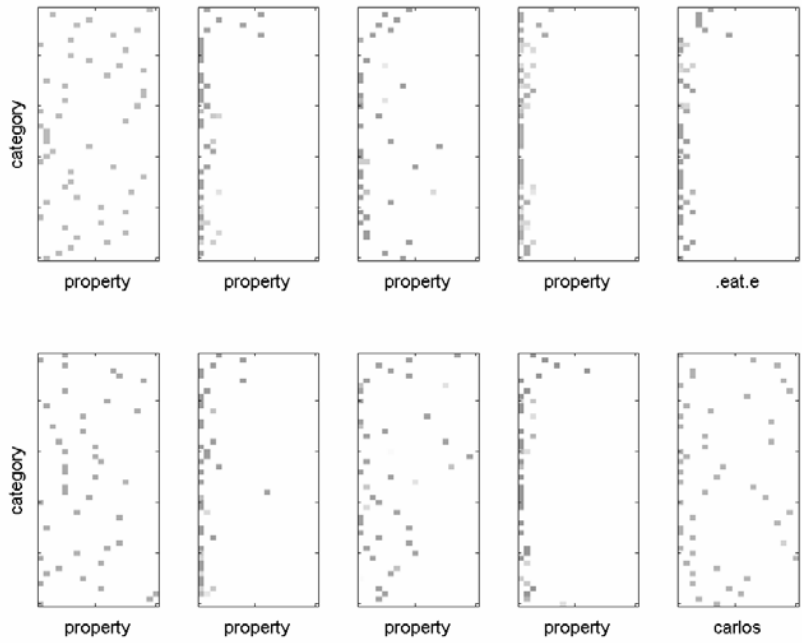

**Figure 8**. Memory retrievals from 6 items of correct memory, taken from the first 10 nominal memories of Fig. 6. Four of the memories are correctly retrieved; the other six attempts at retrieval resulted in junk states rather than the expected memories.

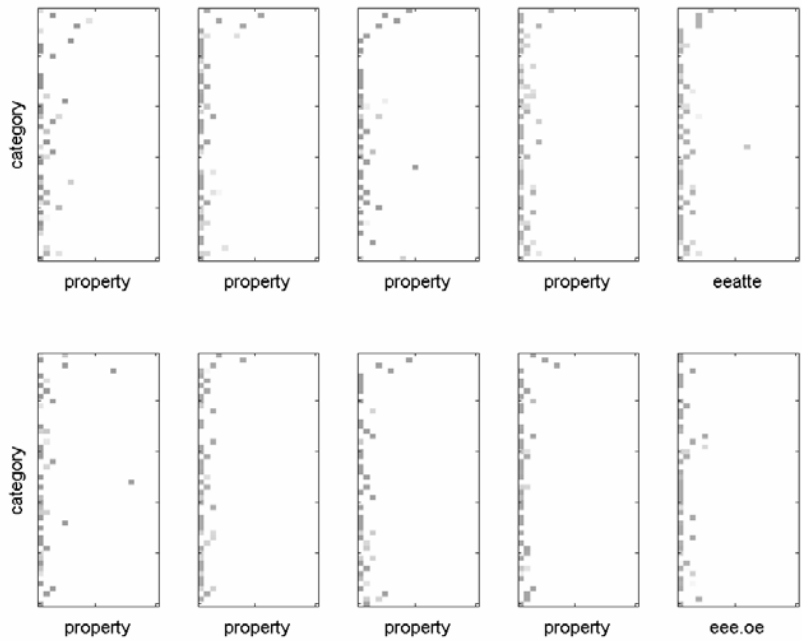

**Figure 9**. Memory retrievals from 4 items of correct memory, taken from the first 10 nominal memories of Fig. 6. None of the memories are correctly retrieved.

By the time that the initial data is impoverished to the extent of having only 6 pieces of information, the retrieval success has dropped to 40%; with 4 pieces of correct information it is zero. Introducing an $i^{ex}_{thresh}$ does not solve the problem. At this level of input information, either there is no retrieval of anything (for large $i^{ex}_{thresh}$) or there is retrieval of states that are mostly junk. At the same time, an ideal algorithm would succeed in finding and reconstructing the correct memory. The overlap between the initial state and the real memories shows this to be the case, as is illustrated in Fig. 10.

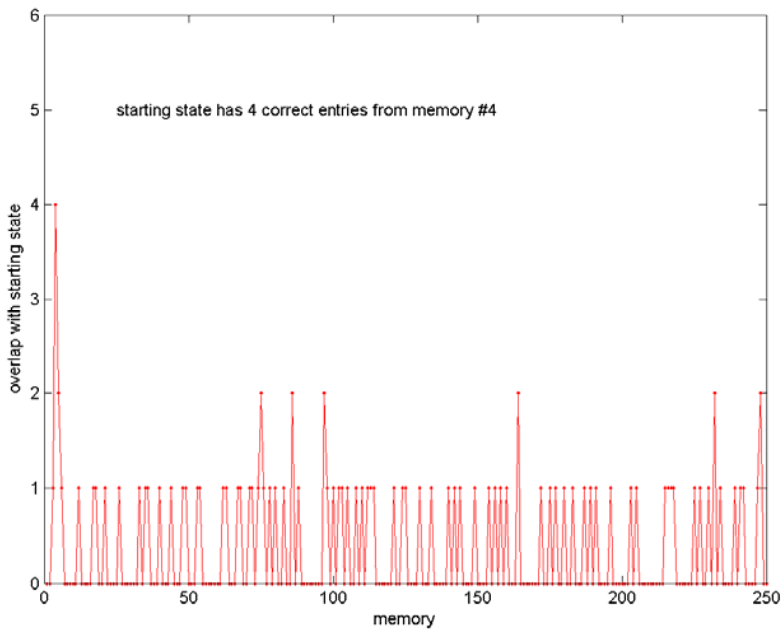

**Figure 10**. The overlap between four pieces of information taken from state 4 and the 250 nominal memories. The maximum possible overlap is 4.

There is overlap 4 only with the memory from which these 4 items of starting information were chosen. Thus these clues provide sufficient information for an ideal algorithm to retrieve the correct memory. The same is true for each of the 250 memories. The fact that the associative memory fails to retrieve correct answers from an adequate clue displays the non-ideality of the retrieval algorithm actually carried out by the dynamics of the network.

These results were obtained by giving the known subset of the components of u non-zero initial values and then allowing all components to evolve in time. This mode of operation would allow for the correction of part of the initial information when there are initial errors. An alternative mode of operation, clamping the known input information during the convergence of all other components, gives similar results.

The problem with recall from weak clues originates from the fact that only excitatory synapses have information about the memories, and that the memories are intrinsically correlated because of the positive-only representation of information, in conjunction with the general skewing of the probability distribution within each category. The result is a T matrix that has many stable junk states, easily accessed while attempting to recover a valid memory. It is the price paid for having a simple learning rule that can add one memory at a time, for working with finite N, for not having grandmother neurons devoted to single memories, and for accepting the expected excitatory-inhibitory discriminations of neurobiology. The benefits to biology of using this less than ideal algorithm include computational speed, Hebbian synaptic single trial learning, and circuit simplicity.

*Can a correct memory be found from a very small clue?*
Suppose we have at our disposal such a dynamical associative memory. Is there any way that we can elicit a correct memory when we know that the cue is actually large enough, but only a junk state comes forth? Can we somehow probe the memory circuit to elicit the correct memory?

One way to search is to insert additional information into the starting state. Suppose we insert an additional 1 in the starting state, in a category where we do not know the correct property, placing the 1 at an arbitrary position along the 'property' axis. If this is not a correct value for the memory being sought, the network should still retrieve a 'junk' state. If it is the correct value for the memory being sought, then the state retrieved should either be a junk state that more nearly resembles the actual memory, or even the true memory state itself.

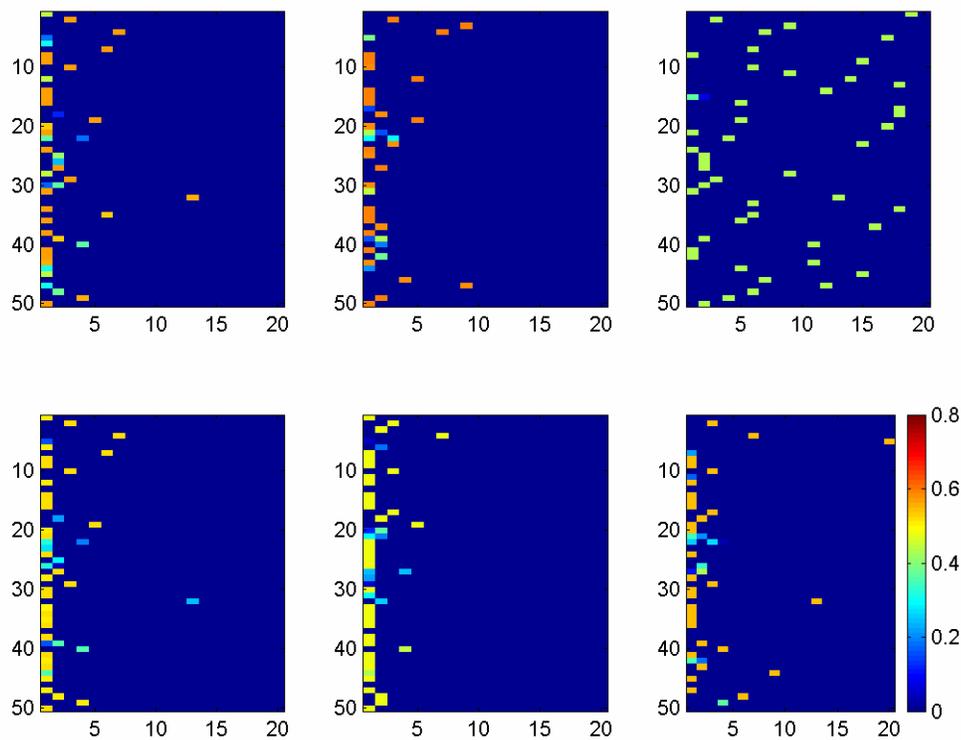

**Figure 11**. Memory retrievals from 4 items of correct memory augmented by the arbitrary assumption that one additional piece of information is correct. Six such examples shown. When that additional piece of information is in fact correct, the system converges to *the* correct memory (top right) and the fact that the memory is *a* valid memory is pop-out obvious. Display as before, but in false color rather than gray scale.

Results from this procedure are illustrated in Fig 11 for the case of the first memory, with a correct cue in the first four categories. This starting cue of four entries returns a junk state. When, however, a '1' is tried in each of the 20 positions of the $5^{th}$ category, 19 times a junk state results; one time the correct memory results and is immediately obvious. Six examples of this retrieval process are shown in false color in Fig. 11. The correct retrieval is clear from its general pattern, a typical memory pattern and having all activities the same rather than a typical junk pattern. Overall, there are 920 unknown positions in which a 1 might be tried; and 32 of them return the correct memory state.

These memory patterns contain intrinsic 'check-sum' information. The existence of a clear categorical difference (that could be quantified and described in mathematical terms) between 'memory states' and 'junk states' makes it possible to identify patterns that are true memories. We can thus retrieve a memory state by querying the memory multiple times, using guessed augmentation, when the initial clue is sufficiently large to identify the memory in principle, but is too small to result in single-shot retrieval because the memory system intrinsically incorporates a less-than-ideal algorithm. After each query the memory returns a state, usually junk, but junk states can be identified by inspection. The number of queries necessary, even when working at random, is far less than the number of memories. The hit rate can be improved by using intelligent ways to select the trial additional information.

*False color in Fig. 11*
To explore what happens in our visual perception, false color was used in Fig. 11. It somewhat enhances the pop-out recognition of correct and incorrect patterns, but the pop-out is strong even in a grey scale representation. One might think of false color as a cheat—an artificial re-representation of the input analog information so that more of our visual processing can be brought to bear. However, the transformation of one analog number into three RGB values is a trivial computation that can easily be given a neural basis. Given that fact, the pop-out so visually apparent to us in the earlier gray-scale pictures might actually involve the generation of a higher dimensional signal in the brain, like the pseudocolor representation of intensity, as part of our neural computation.

*Aha-correct! = pop-out*
The effort to retrieve memories from weak clues inevitably leads to the frequent retrieval of patterns that do not represent true memories. Because uniformity of intensity, uniformity of color, and spatial pattern balance are all pop-out discriminators, real memories (regardless of length) and erroneous or junk results of a recall effort have patterns of activity that are visually distinguishable in a pop-out fashion. Therefore the same neural circuit wiring and dynamics that produce visual pop-out will also provide the ability for rapid differentiation between instances when a retrieved memory is valid, and when the retrieval has not resulted in a valid memory. While there may not yet be a complete theory of visual pop-out, we know from this similarity of computational problem that higher areas will be able to rapidly discriminate this distinction. The strong feeling that a retrieved memory is correct is computationally isomorphic to pop-out.

*Complex patterns also permit the recognition of inaccurate recall*

For this memory model, unequal analog values of the activities of the neurons in the final state reached in the retrieval process are an indication of an erroneous state. The failure to have an appropriate distribution across the horizontal dimension also permit this recognition. When the memories have additional statistical structure, there are additional mechanisms available. We illustrate such a case.

In 2 of the 250 memories, the lowest 6 rows of the state matrix have been used to encode 20 of the letters in the alphabet, in order of probability of usage in English. For these two memories, the names 'jessie' and 'carlos' were written into these 6 positions so that the decode of the correct memory states would produce those names. In the Figs. 5-9, the horizontal axis is always 'property'. It is labeled as such *except for these two memories* (located at the right hand side of Figs. 5-9 ) where instead of 'property' the recalled reconstruction of the name is given. When the correct memory is recalled, as in Fig. 6, the names are correctly given. When, as in Fig. 7 and Fig 9, there is a major problem in the recall process, the collection of letters that is returned is obviously not a name. In this case, the junk states returned contain only vowels, and what we quickly observe is due to the fact that the first order statistics of the letters does not correspond to those in names.

But consider the sequences 'wscnot' 'alanej' ' ttrdih' 'dtwttj' hwljpe' 'heeain' 'hpoioi' 'cantpo' which are random uncorrelated letter sequences generated using the actual English letter-frequency distribution. Although these sequences have the correct first-order letter statistics, the sequences are recognizably non-names (except for the last) because they violate the spelling rules of English. Thus a higher area that knows something of the spelling rules of English, but has no knowledge of particular words, could with fair reliability tell the difference between a case of correct recall and a junk memory even if the junk memory has the correct first order statistics.

It is often argued that information should be transformed before storage in such a way that the transformed items to be stored are statistically independent. This is an effective way to reduce the number of bits that must be stored. The transformed items actually stored appear random. However, removing all correlations and higher order statistical dependences eliminates the possibility of examining whether the results of a memory query are correct or have been corrupted, since any returned answer is equally likely to be a valid response. By contrast, if some correlations are left, then a higher area making a memory query has a means by which to evaluate the quality of the returned information.

*Parallel search of larger memories*

Searching many memory modules can readily be carried out in parallel. Fig. 12 shows the basic multi-module circuit. There are many modules, which anatomically might correspond to many hypercolumns. While the present discussion will treat these as separate and delineated (like barrel cortex in rats), the basic ideas described here should be extendable to a cortex which is merely periodic and less artificially modularized. Inputs come to corresponding cells in different modules via branched axons from another brain area. Outputs from corresponding cells in different modules are summed by pool cells. When the clue to the memory is substantial and the threshold parameter $i^{ex}_{thresh}$ is properly set, a module queried with the clue will find the appropriate memory if it is contained within the module, and will go to the all-off state otherwise, as shown earlier. If the outputs from corresponding cells in all the modules are summed by pool cells (see figure below), the output of the pool cells will be due to the correct memory found by one of the modules, since all other modules will have no output. This search is fast, requiring only the time necessary for a single module to converge. The all-off state of each module provides an automatic choice of the correct memory module to drive the pool cells.

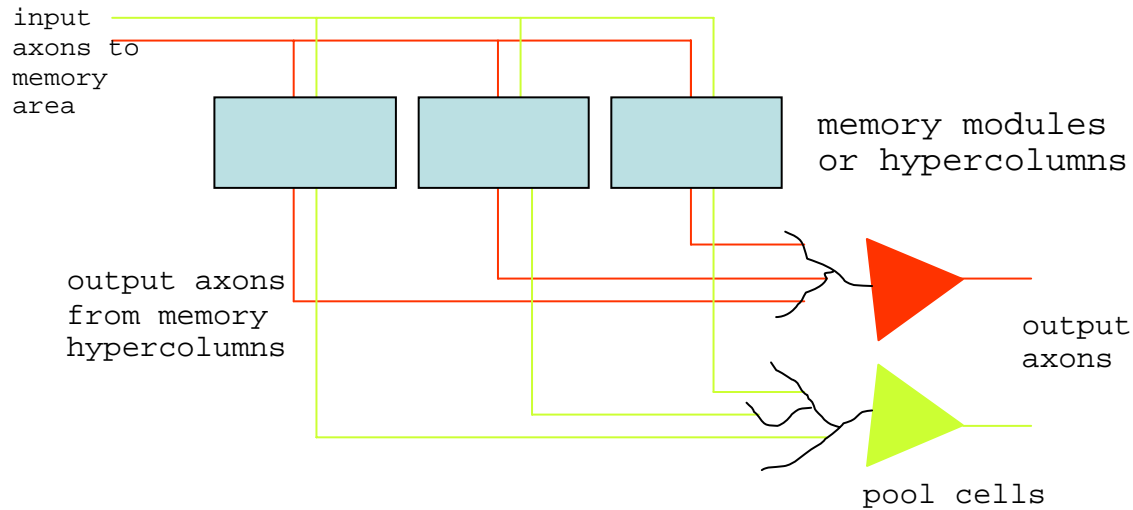

**Figure 12**. Equivalent memory modules (blue squares) all receive the same inputs from a set of axons carrying the query information. Equivalent cells in each module send signals to the same pool cell. If only a single module remains active after the query, the pool cells will display the activity of that module.

*Parallel search and synchrony in the case of weak memory clues*
When the clue is rather small in size but still adequate for reliable retrieval, $i^{ex}_{thresh}$ must be set so low that this simple procedure begins to fail in a multimodule system. Another module that does not contain the correct memory can settle into a junk state of activity, and the input to the pool cells would consist of the correct memory from one module summed with the junk state(s) of another module(s). One conceptual possibility is that each module explicitly evaluates whether it has found a true memory or not, and sends signals to the pool cells only if it has found a true memory. Any computation that pop-out recognizes a true memory could be used to gate the output from a module. Such a method of operation would require neural circuitry to explicitly gate each module. A more plausible possibility is to make use of spike timing and synchrony, which could achieve the same result without the need for additional circuitry or explicit gating.

A group of spiking neurons that have the same firing rate and are weakly coupled often synchronize. This synchrony can be a basis for the many-are-equal (MAE) operation, a recognition that a set of neurons is similarly driven and not diverse in firing frequencies. MAE was usefully implemented for speech processing [Hopfield 2000; Brody 2001]. The difference between the recollection of a correct memory and a junk state is precisely that the correct memory has the firing rates of all neurons the same. We therefore expect that if spiking neurons and appropriate synaptic couplings are used instead of using a rate model, a correct recall will result in synchrony of the active model neurons, while a junk state would not result in synchrony. When there is only a single module, synchrony can be used as a means to detect correct versus junk states when the query information is weak. Here, we discuss the extension of this idea to the problem of retrieval of a memory from a multimodule system when the query information is weak, so that modules may respond individually with the correct memory or with a junk state.

The model description so far described has one 'neural unit' for each point in the category-property grid. Neurobiology would undoubtedly use more. Suppose that we increase this number to 10 for each point, leaving all else the same. These 10 will all get common input signals, and will all send their axons to the same pool cell, and have the same pattern of synaptic interconnection that the smaller model did. Increasing the number of neurons will increase robustness, but in addition has new encoding capabilities.

In a junk state, the 10 outputs corresponding to a particular category and property will fire at the same mean rate, but because the module does not synchronize, the timing of the action potentials of the 10 neurons will be arbitrary. In a true memory state, again these 10 neurons will fire at the same rate, but because the module will synchronize, these 10 neurons are now well-synchronized. If the pool cell connected to these 10 neurons has a short time constant, it will fire in response to the latter circumstance, but not the former. Thus the pool cell will have an output only when a true memory has been found as a result of a memory query.

Now let many of these enlarged modules be all connected to the same pool cells. After a memory query, the module that contains the correct memory will synchronize, while the others either go to a junk non-synchronizing state or turn off. The input from junk non-synchronized states is ineffective in driving a short time-constant pool cell, so the dominant pool cell response is a volley of spikes *at the firing rate of the synchronized module*. The pool cell has implicitly identified the module with the correct information, and in its spiking response will chiefly disregard the other inputs to it. This magical-sounding result of being simultaneously driven by multiple modules but responding only to the module with the correct answer is a consequence of local synchrony and the implicit meaning of such synchrony.

In concept, the pool cells are common to many modules, and may be in a separate anatomical region or brain area. A hallmark of such processing is a transient local quasi-periodic EEG in one location α (that contains the memory), which in turn induces a transient local EEG in another location β (that is the common readout pathway for many modules) and is phase-locked to the transient in α. A variant of the same task invoking a different memory module would lead to an EEG transient in location α' without one at α, which induces one at β that is now coherent with the transient at α'.

*Error-correcting codes*

Digital error-correcting codes lead to the ability not just to recognize when errors are made, but also to correct the errors. Suppose that we are sending blocks of N bits, so the possible transmitted signals are the corners of an N-dimensional cube. However, instead of allowing any of the $2^N$ corners to be sent, we will restrict ourselves to choosing data to be sent from a smaller subset of $2^P$ corners. This subset can be chosen so that they are well separated. Each member $X_k$ of this subset is surrounded by a region of $2^{N-P}$ nearby corners that are never deliberately transmitted, and which are nearer to $X_k$ than to any other $X_{k'}$.

The receiver of the information knows what codewords are allowed to be sent. If the receiver receives an allowed codeword, it is presumed to be correct. If the receiver finds a non-codeword, it is replaced by the allowed codeword that is closest to it. This decoding process corrects errors and restores the information originally sent. As long as restoration is done before too many errors accumulate, the process can be repeated over and over and can completely prevent information loss. We will show that a related procedure can be used for keeping our memory error-free against synapse drift, by occasionally evoking each memory and correcting errors while they are still small.

*Check-bits, error correcting codes, and the long-term maintenance of memories*

The long-term maintenance of memories that are stored in synapses presents a conceptual problem. The literal use of error-correcting codes, as in computer science, is out of the question. Storage of the same memory independently at multiple places in the brain is a possible aid to memory maintenance, for 3 copies can be occasionally intercompared, and if one of the copies has changed it could be rewritten to agree with the other two. This seems more a digital engineering solution than one that biology will exploit, for it is both wasteful of storage space and requires a very logical form of intercomparison.

The problem of long-term memory storage is somewhat alleviated in the present modeling due to the use of binary synapses. However, individual synapses might well be binary with somewhat different values of 'on' when they are made  Or these values might themselves drift. Synapses might be lost, even if binary. The system will need a means of recognizing and compensating for such problems if it is to function well over long time periods, and if the uniform activity characteristic of a proper memory is to be maintained. We now show in concept how spike-timing-dependent plasticity can act to restore correct memories and eliminate memory drift, *exploiting the fact that correct memories are distinguishable as correct*, and utilizing the phased synchrony of coupled neurons that are almost equally driven.

Our visual system quickly picks up the differences between the three patterns in Fig. 13, which represent a spurious memory; an ideal memory; and an ideal memory for which all non-zero synapses (which should all have the same value) have additive Gaussian noise. The central pattern is in need of synaptic adjustment to make the firing rates of the active neurons equal again, while leaving inactive neurons inactive. A similar figure can be generated when noise is due to synapse loss.

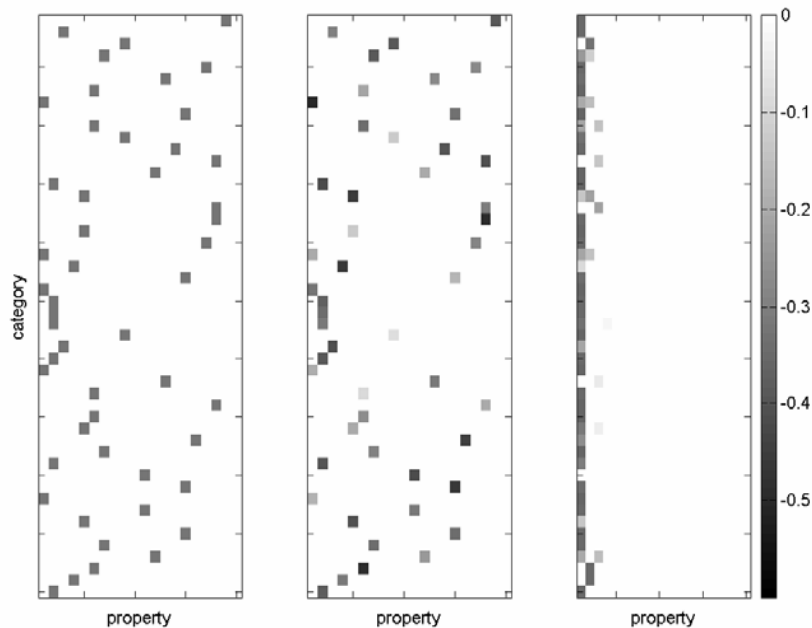

**Figure 13**. Activity patterns for a spurious memory (right) an ideal memory (left) and an ideal memory when all non-zero synapses have Gaussian noise (center) with σ = 5%. The degree of difference between the activity of the neurons in the central panel has been reduced in the simulations of this particular figure by adding some negative curvature in the input-output relation of excitatory cells (see Fig. 4), which increases the dynamical range of synaptic errors over which the activity pattern is still identifiably related to the corresponding memory. Without this curvature, a noise level of 2% would generate the level of differences seen in the central panel. The steepness of the inhibitory input-output function also affects this sensitivity. No exploration of the relevant parameter space has been made.

If cell and synapse physiology provide a mechanism for making a synaptic correction toward equal firing rates whenever such a pattern of excitation is found, valid memories will be able to combat random synapse drift or random resetting and remain valid memories. In concept, the system can behave as if there were an error-correcting code. Proper memory states have the same activity strength for all active neurons, and values of the synapses that will result in that sameness. Each proper memory state with proper synapses is surrounded by a set of almost proper states with slightly incorrect synaptic connections (resulting in unequal activity for this same set of neurons due to the incorrect values of synapses). Other proper memory states are far away, with different neurons active. This configuration is in concept like that of the engineered encoding for error correction described in the previous section. What is needed is a restoration process that will make synaptic changes convert an almost proper state into the nearby proper state.

Consider the effect of synchrony in an implementation with spiking neurons. When the network is started from random initial input, the network may converge to a junk state which will not result in synchronization. A spike-timing plasticity rule of the type often seen experimentally [e.g. Bi 2001] , with synaptic potentiation accompanying pre-before-post synaptic spike pairings, and depression accompanying post-before-pre spike pairings, will not result in synapse change for uncoordinated pre- and post-synaptic action potentials.  If the system converges to a correct memory and all the neurons are driven the same way, synchrony should be precise, which can then result in no synaptic change due to the location of a zero in the spike-timing-dependent learning rule, which changes sign near zero pre-post delay.   If the synapses have drifted such that a correct memory results but the neurons do not all have quite the same input, then although all neurons may be firing at the same rate, some will lead and some will lag in their timing. The usual spike timing dependent synapse plasticity behavior will then result in synapse change based on this fact, with synapses to cells that have somewhat weak total inputs being strengthened (or changed from 0 to 1 in the case of a binary system) while synapses to cells with total inputs somewhat too large will be weakened (or changed from 1 to 0 in the case of a binary system).  The net result will be to weaken the drive to neurons that are too active, strengthen the drive to neurons that are not active enough, and move the activity pattern toward the ideal pattern for that memory.

This example provides a conceptual framework in which to understand long-term memory maintenance.  When memories can, in their pattern of activity, be identified as 'correct', and valid memories are appropriately isolated, memory rehearsal should drive spike-timing-dependent synaptic plasticity of the usual type in such a way that it will maintain the pattern of the memory activity states in the presence of synaptic birth, death, and drift.  The synaptic patterns themselves may drift when this procedure is followed, but the memory activity states should not drift, except for a possible (and meaningless) scaling drift of the level at which all cells are active in a valid memory.

**Sudoku puzzles**

Sudoku puzzles [Hayes 2006] are written on a 9 x 9 board.  A particular puzzle is described by digits in the range 1-9 written in a few of the squares.  Fig. 14 shows a typical 'fiendish' puzzle taken from the *Times Su Doku Puzzles 1* [Gould 2005].

|   | 1 |   |   |   | 8 | 4 |   | 7 |
|---|---|---|---|---|---|---|---|---|
| 9 | 5 |   |   |   |   |   |   |   |
|   |   | 8 |   | 1 |   |   |   |   |
|   | 8 | 2 |   |   |   |   |   |   |
| 7 |   |   | 4 |   | 6 |   |   | 8 |
|   |   |   |   |   |   | 6 | 2 |   |
|   |   |   |   | 5 |   | 7 |   |   |
|   |   |   |   |   |   |   | 8 | 2 |
| 5 |   | 3 | 2 |   |   |   | 1 |   |

**Figure 14**.  A difficult Sudoku puzzle. The problem is to fill in the blanks with entries 1-9 such that each digit occurs once in each row, once in each column, and once in each of the 9   3 x 3 squares indicated by double lines.

Each puzzle has a unique solution.  Very simple puzzles can be solved using elementary logic, such as noting that if a blank space has eight different digits in its surround of its row, column, and 3x3 square, then that blank must contain the other digit.  More difficult puzzles require more complex logic.  For very difficult puzzles most people reach a point in the solution process at which they make an intelligent guess about a new entry in the matrix, and follow the consequences of that guess to the solution (or to a demonstrable inconsistency, then backtrack to the guessed entry, and guess differently … ).  While the composers of puzzles sometimes remark that in their design each is solvable by a logical procedure and guessing is not needed, the degree to which logic must project forward into a sequence of steps is left undefined.  (Even exhaustive search is a logical procedure, but of a rather lengthy logical chain.)  The most difficult Sudoku puzzles, generalized to $(3^2)$ x $(3^2)$  to $(n^2)$ x $(n^2)$ are NP complete problems [Yato 2002, Hayes 2006].  We consider only the usual 9 x 9 puzzles.

*A 'neural' representation of the puzzle*

A puzzle is a matrix $X(i,j)$. A puzzle is defined by $X(i,j) = 0$ except for the given entries, e.g. $X(1,2) = 1$, $X(1,6) = 8$, etc. in the puzzle above. A finished puzzle has all $X(i,j)$ integers 1-9 obeying the rules described. To construct a 'neural network' Sudoku solver, rewrite the problem in a binary form in terms of a 9 x9 x 9 matrix $V(i,j,k)$ with entries 0 or 1. The correspondence between V and X is given by

$$\text{if } X(i,j) = 0, V(i,j,k) = 0 \text{ for all k.} \qquad (10)$$
$$\text{if } X(i,j) = k \neq 0 , V(i,j,k) = 1$$

For a given k, the $(i,j)$ locations of the 1's in $V(i,j,k)$ describe the locations of the digit k. Thus in the V notation, the initial board above is $V(1,2,1) = 1$; $V(1,6,8) = 1$, etc. In the binary representation, the rules for the solution of Sudoku become

$V(i,j,k) = 1$ for the set locations defining the particular puzzle.

$V(i,j,k) = 0$ or 1 for all i,j,k

$\sum_i V(i,j,k) = 1$ for all j,k $\qquad (11)$

$\sum_j V(i,j,k) = 1$ for all i,k

$\sum_k V(i,j,k) = 1$ for all i,j

$\sum_{i,j} V(i,j,k) = 1$ for all k, with the sum on i and j taken over one of the 3x3 i,j squares bounded by the double lines.

There are a total of 4 x 81 = 324 constraints on sums of elements of V. The constraints allow a maximum of 81 1's to be put into the matrix, describing the 81 integers of $X(i,j)$ in the solution.

The Sudoku problem in the V variables can be posed as the 'integer programming' problem

Maximize $E = \sum_{i,j,k} V(i,j,k)$ subject to the above constraints. $\qquad (12)$

*The obvious neural approach does not work well*

With this description as an optimization problem on a set of binary variables, neural networks can be designed to try to solve a puzzle. Because Sudoku requires a binary answer, the most obvious approach is related to the way in which the Traveling Salesman Problem was posed on a neural network [Hopfield 1985]. Let each V be represented by a sigmoid model 'neuron' with minimum output 0 and maximum output 1. Define the domain of a neuron $(i,j,k)$ as all the other neurons that have the same i, the same j, the same k, or are in the same 3x3 square (at the same k) as is neuron $(i,j,k)$. The inhibitory pattern is reciprocal from the rules of the puzzle. We wish to maximize E (above) with the constraint that each neuron that is 'on' (with an output of 1) has all other neurons in its domain 'off' (0 output). Define the Energy function

$$\text{Energy} \;=\; -\sum_{i,j,k} V(i,j,k) \;+\; \alpha \sum_{i,j,k,i',j',k'} V(i,j,k)\, V(i',j',k') \qquad (13)$$

where the sum is taken over all neuron pairs within a domain. This energy function defines the necessary connections in a feedback neural network solver. If the gain of the sigmoid neurons is large, in steady state each neuron will be either 'off' or 'on'.

The information about a particular puzzle can be inserted by holding the appropriate subset of neurons 'on'. The system begins its dynamical motion from the state with all other neurons 'off'. The dynamics of the neuron activity variables leads to an attractor state that is a local minimum of Energy. The desired solution is the global minimum of Energy. Since we are given that there is a unique answer, and at this answer the inhibitory terms are all zero, we know that the desired solution will be the lowest minimum of the Energy function. There is no guarantee that the neural dynamics will find the true minimum--it may instead find only a local minimum. When tried on puzzles taken from *The Times Su Doku Book 1*, this network finds the correct answer to only the simplest examples. In more difficult problems the dynamics converges only to a local minimum that is not the desired solution.

*A neural network coprocessor for Sudoku*
The network of the previous section attempts to solve the computational problem in one convergence step. When this process fails, there is little in the answer to suggest what to try next. A network that is to be used as a fast coprocessor should, if it fails to provide an exact answer, at least provide a limited range of possibilities to be considered by the next processing stage. That next processing stage can then perform a different sort of computation on the basis of the current inputs from the coprocessor, or it may return additional information to the coprocessor, asking the coprocessor to compute again on the basis of the additional information.

Neural networks can exactly solve the 'linear programming' problem [Tank 1986], maximizing a linear function (like E) with a set of linear *inequality* constraints on a set of *continuous* variables. Such a neural network solves the problem through its state evolving (over time) from an initial condition describing the problem to a final stable state in which the answer can be read out from the activity level of the neurons. Inspired by the fact that some integer programming problems can be solved in a space of continuous variables [Smith 1996], we construct a linear programming network to solve the following problem

$$\text{Maximize} \quad E = \sum_{i,j,k} V(i,j,k) \qquad (14)$$

on continuous variables $0 \leq V(i,j,k)$ with 324 linear constraints obtained by replacing the '=' in the integer programming version by '$\leq$'. The network contains 729 'V' neurons, with a semi-linear response, and 324 inhibitory constraint neurons, one for each inequality constraint. These constraint neurons have an input-output relation that is zero when their input is less than unity, and rises very rapidly when the input rises above unity. Each inhibitory neuron receives a synapse of unit strength from each of the neurons appearing in the inequality represented by the inhibitory neuron, and in turn makes equal inhibitor connections onto this same set of V neurons.   In addition, each particular puzzle is represented by holding 'on' (value 1) the subset of V described by a particular puzzle.

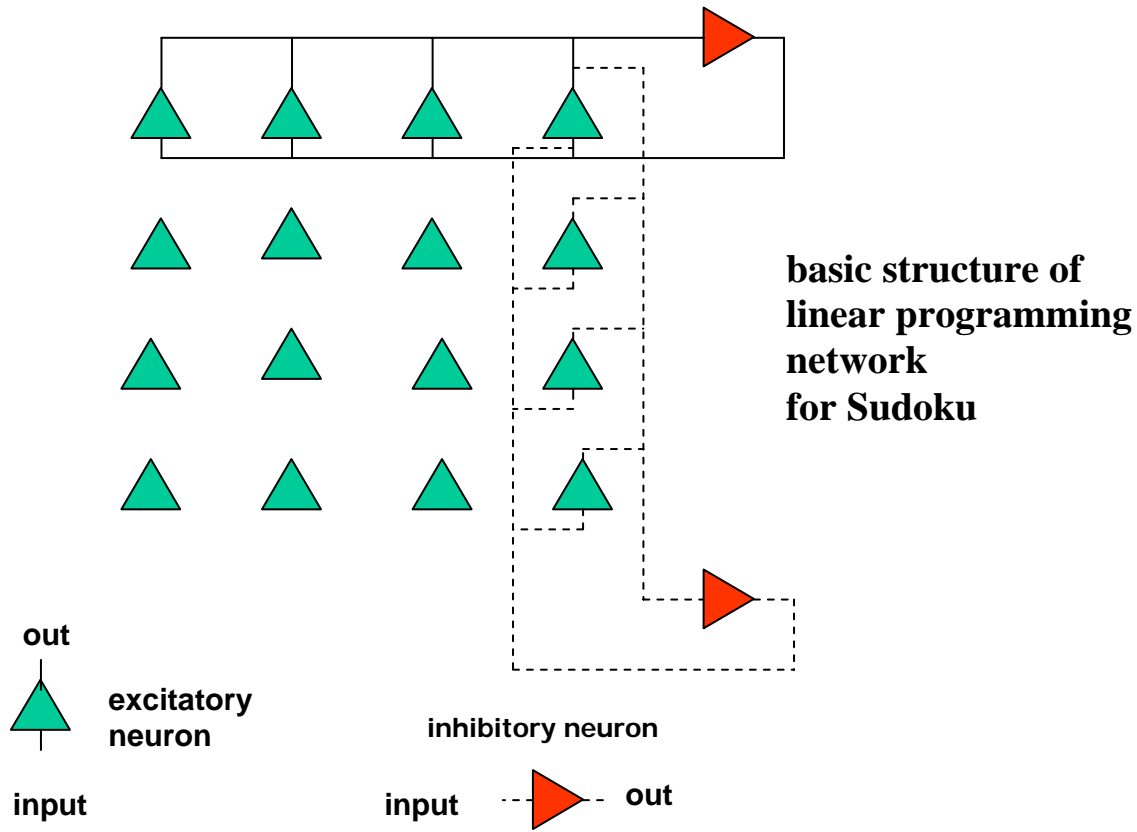

**basic structure of linear programming network for Sudoku**

**Figure 15**. The network that solves the linear programming problem [Tank 1985] for Sudoku.  All model neurons are of the linear threshold type.  There is an inhibitory neuron for each constraint, with a threshold set so that an inhibitory neuron turns on strongly if and only if its constraint is violated.  All excitatory neurons are equally driven, and  are equally coupled to each relevant constraint neuron.

What should we expect to result from a convergence of this linear programming neural network to a solution?   We know in the present case that there exists a solution with all V's 1's and 0's, that it has a value of E = 81, and that no points within or on the surface of the convex polyhedron represented by the inequalities have a greater value of E.  The desired solution is at a vertex of the polyhedron.  There are no other vertices having all values of V= 0 or 1 that also have E = 81, for the solution is unique.  However, there might be other vertices with values of V lying *between* 0 and 1 that also have E=81.  Let us call these 'illegal solutions'.  If there are no illegal solutions to a particular puzzle, the coprocessor network will find the correct answer in a single convergence, a uniquely maximal vertex of the polygon, and will converge to this answer independent of starting point.  If there are N independent illegal solutions, there exists an N dimensional hyperplane subspace on which E = 81.  This hyperplane of has a polygonal boundary, and the desired solution is at one corner of this polygon.  Unfortunately the neural network dynamic trajectory of V will in this case continue evolving only until V hits this subspace, and then V ceases to change. The location at which V stops then depends on the initial values of the variables, is not unique, and is not the desired solution.

*The performance of the 'coprocessor'*
The convergence of the neural net coprocessor was examined for Sudoku puzzles from three sources. The source matters, for these puzzles are themselves generated by programs, and the methods of generation do not provide random examples of all possible puzzles. Applied to 20 Sudoku puzzles taken from the *Times Su Doku Book 1*, the processor solved all of them correctly, including 10 called 'fiendish'(highest level of difficulty). Figs. 16-19 show the convergence of V to the solution of one of these puzzles.

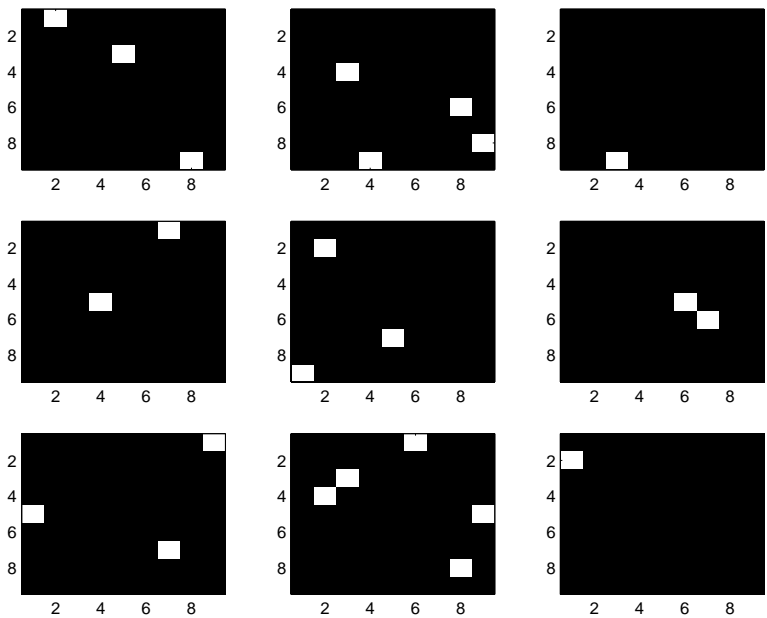

**Figure 16.** The values of V for the **initial state** (Fig 14 )of the network. Each of these 9x9 arrays represent a 'k' plane of V(i,j,k). Black is a zero, white is a 1. The top left display has 1's in locations (1,2), (3,5) and (9,8) representing the fact that the initial state has 1's in locations (1,2), (3,5), and (9,8). The top center display has 1's in locations (4,3), (5,8), (8,9) and (9,4) representing the fact that Fig. 14 has 2's in these locations. The top right display represents the fact that Fig 14 has a single 3, at location (9,3).

All the bright squares in this and subsequent figures should be the same size. While this is the case for the screen display of this pdf, it is not the case for some printers due to printer driver errors. The .doc version of this paper, available at http://genomics.princeton.edu/hopfield/, can be more printer-friendly.

Throughout the convergence, entries of V that pose the puzzle are fixed at 1. The correspondence with the earlier figure can be seen in these panels. For example, the bottom right panel indicates that in the puzzle to be solved, the only known location of a 9 is the first entry in the second row.

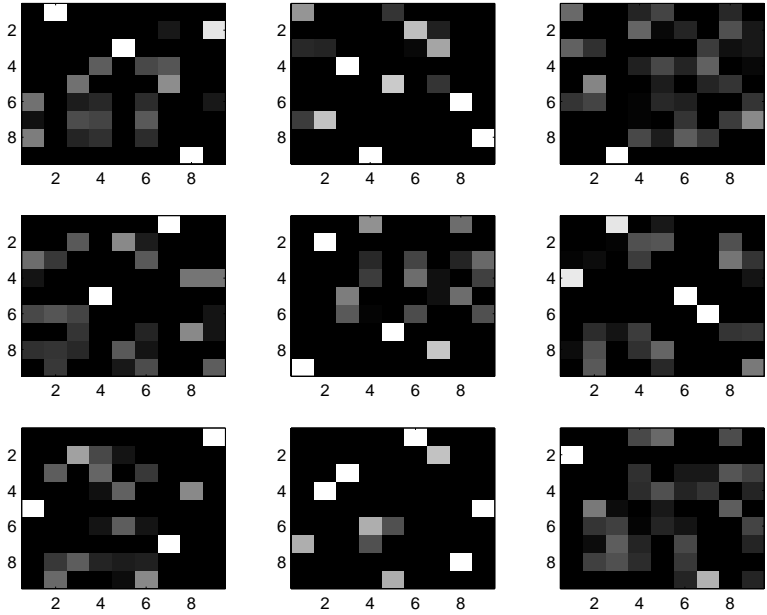

**Figure 17**. The values of V for an *early intermediate time during a convergence* of the network when started with the initial state of Fig. 16. Each of these 9x9 arrays represent a 'k' plane of V(i,j,k) as in Fig. 16. Gray scale indicating the values, with black as 0, white is a 1.

Initially, all elements of V except those given in posing the puzzle are set to zero. In the intermediate states, other elements turn partially on and take values between 0 and 1. Any (i,j,k) can be thought of in terms of the proposition that integer k should be located at position i,j. If the proposition is true, V(i,j,k)=1; if false, 0. These intermediate states reflect partial 'belief' in often contradictory propositions. Ultimately these contradictions are resolved, and the network obtains the steady state V below.

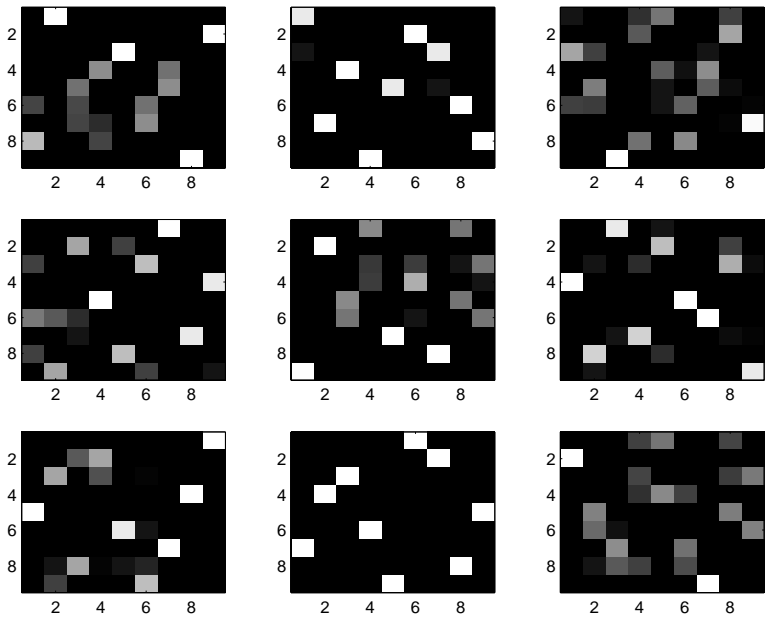

**Figure 18**. The values of V for a *later intermediate time during a convergence* of the network when started with the initial state of Fig. 16. Display as in Fig. 17.

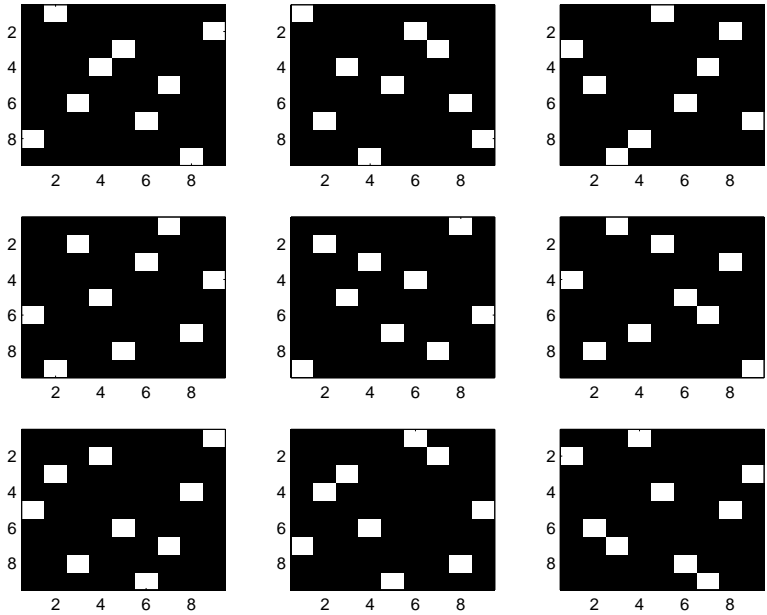

**Figure 19**. The *final state* of V after convergence when started with the initial state of Fig. 16. Display as in Fig. 17. While the display is gray scale, all the actual values are now either 0 or 1.

In this final state, each proposition is either true or false, and the propositions that are true provide the correct answer to the puzzle.

The Daily Telegraph book of Sudoku puzzles [Mepham 2004] contains puzzles ranked as gentle, moderate, tough, and diabolical. Gentle and moderate puzzles are exactly solved in a single convergence. Most of the tough puzzles are solved, but only about half of the diabolical ones are. When no solution is found, the convergence arrives at a point on a high dimensional hyperplane where E = 81 and stops there. The dimensionality of this hyperplane can be determined by starting the system with a random initial bias, then allowing it to converge. When there is a unique solution, it will be found regardless of the initial condition. However, when there is a hyperplane where E = 81, this procedure will generate a 'random' point on the hyperplane. The dimensionality of this hyperplane can be found by starting from many random points and applying PCA. The dimensionality of this hyperplane was determined in 20 cases, and was found to range from 10 to 34.

'Can be done by a logical procedure' has been given a more mathematical description [Eppstein 2006]. David Eppstein provided me with a set of 10 very difficult puzzles that could be solved in his logical fashion, and 10 others that were so difficult that their solution seemed to necessitate backtracking (in his definition). When the linear programming network was applied to each of these examples, V ended at a point on a high-dimensional hyperplane. There was no obvious relationship between difficulty and dimensionality.

*Failure to reach a correct solution has visual pop-out*
A typical final state for a convergence to a high-dimensional subspace for one of Eppstein's examples is shown in Fig. 20.

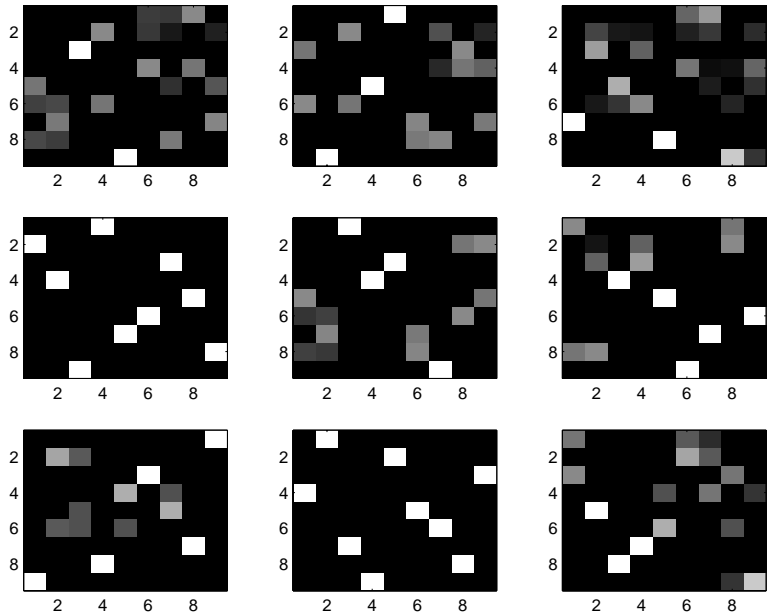

**Figure 20**. The final state of the network after convergence when the puzzle is very difficult and the network fails reaches a final state that violates the expected syntax, and thus cannot be decoded. Display as in Figs 16-19, but the puzzle was very difficult, and not that of Fig. 16.

It is immediately recognizable as a non-solution from the fact that the final firing rates of the neurons are not all the same. And while two of the 9 x 9 squares have unit activity of one neuron in each row and each column, the others instead have fractional activity of more than one neuron in various rows and columns. This particular example has converged to an arbitrary location in a 14-dimensional hyperplane having E = 81.

*Illegal solutions contain information about the correct solution*
Since the subspace of legal + illegal solutions is a hyperplane, points in this space of solutions can be represented as linear combinations of N linearly independent basis functions that obey all the linear inequality constraints. One of the basis functions can be chosen as the actual solution to the problem. The other basis functions obey all inequality constraints, have only positive (or zero) values of V, but fail to have all entries being 1 or 0. A convergence from a random starting point results in a final state that is a linear weighted sum of the N basis functions. The values of the coefficients will depend upon the starting point.

Except for pathological accidents, when a convergence results in a state where a particular row, column, or post of V has a single 1 and the other entries 0, that must be a feature shared by all the N basis functions, and thus be a feature of the desired solution. When a row, column, or post has only two non-zero entries (instead of 1), all the basis functions, including the correct solution, must have entries only in those two places. Thus the correct solution must have its 1 located in one of these two places.

*Guessing leads to a very short logical tree*

Consider the panel in the lower right of Fig. 20, representing the locations of the digit 9 in the solution. From the grey scale image (and confirmed by examining the quantitative values of $V(i,1,9)$) there are only two non-zero entries in the first column, namely (1,1) and (3,1). The correct solution must have a 1 in one of these locations. We might for example guess $V(1,1,9) = 1$ as *an additional fixed and initial condition, as if it had been given in the original puzzle*. With this guess, either

i)      the convergence goes to E = 81, and all entries are 0 or 1, and this is the desired unique solution

ii)     the convergence goes to E < 81 and $V(1,3,9) = 1$ must be true in the desired solution

iii)    the convergence goes to E = 81, but not all entries are 0 or 1.

The guess $V(1,1,9) = 1$ leads to circumstance i), the correct solution shown in Fig. 21 If this guess was made first, only one convergence of the coprocessor network (beyond the initial one) is necessary to obtain the correct solution.

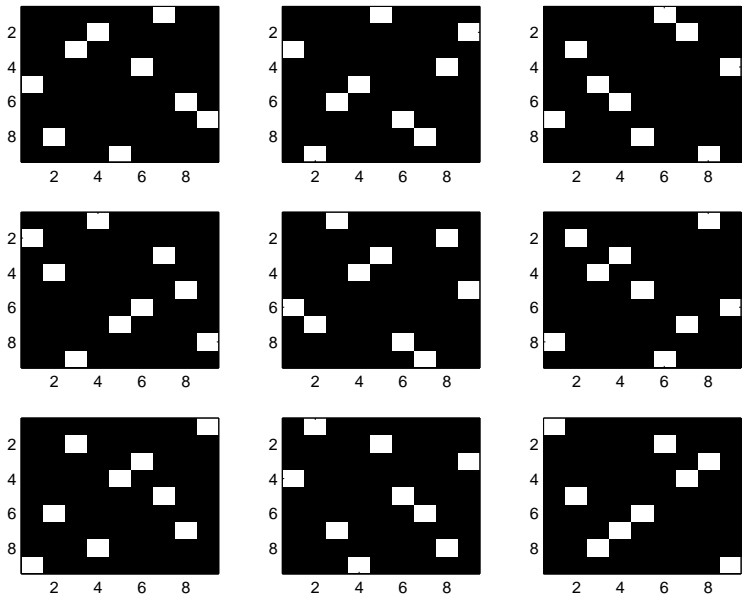

**Figure 21**. The final state of the network for the puzzle used in Fig. 20 when one additional piece of information was given, namely the guess that the correct answer would have guess $V(1,1,9) = 1$, i. e. that the answer would have a 9 in position (1,1). It is visually obvious that the solution has been found.

The alternative guess $V(3,1,9) = 1$ leads to a non-solution (Fig. 22) with $E = 80.71$, circumstance ii). If this guess had been made first, we would now know that the other possibility $V(1,1,9) = 1$ must be correct, but would not yet know the full solution until a convergence was carried out including this new knowledge.

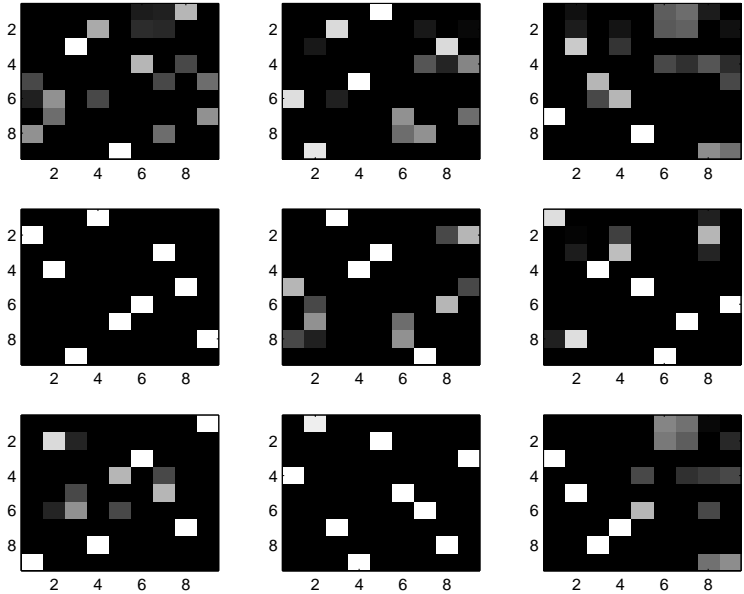

**Figure 22**. The final state of the network for the puzzle used in Fig. 20 when one additional piece of information was given, namely the guess that the correct answer would have guess $V(3,1,9) = 1$, i. e. that the answer would have a 9 in position (3,1). It is visually obvious that no solution has been found.

Guessing a single entry is not always adequate, but the all the difficult puzzles I have studied in this way lead to logical trees that are only 1 or 2 plies deep.

*Special purpose coprocessors*
Suppose that a hardware linear-programming circuit is cheap, effective, and fast. A very simple logical system can be set up that calls the linear programming hardware co-processor several times in succession, with different inputs, in working its way through a logical tree on its way to the solution of a Sudoku puzzle. The bulk of the computation is done by the coprocessor; the logical analysis that must be done to use this coprocessor effectively is elementary. Guessing based on the appearance of the non-solution convergences solves most of the problems rapidly, though it can be helpful to also use the value of E to reduce the number of guesses necessary.

The Sudoku example illustrates that a neural-type coprocessor which is effective in computation (i.e., requires little neural hardware and converges rapidly to an answer), but intrinsically unable to solve the desired logical problem in general (though it may be able to rapidly solve very simple examples) can take up the major computational burden of finding the solution to difficult problems when called iteratively by a simple processing agent capable of rather limited logical analysis. Complex visual scenes, auditory scenes, and the analysis of language are among the candidates for such iterative computation.

**Discussion**

The ability of these minimal networks to solve non-trivial problems in a single convergence indicates that the design is a highly effective use of neural hardware. However, in difficult cases these attractor networks often failed to produce a correct answer, even though the problem posed has a unique correct answer. We have exhibited models in which a neural system is capable of recognizing when answers are erroneous through the use of 'implicit check-bits'. These are closely related to the concept of check-bits in digital information transmission, but are implicit rather than explicit. The *idea* that the intrinsic check-bits might be used to evaluate the correctness of neural computations transcends the very simple neural circuitry and patterns used in this essay.

This recognition enables the computational power of a module to be more fully exploited through sequential use of the module by the larger system. In the present examples the signals flowing from higher centers to the lower center are equivalent to additional sensory information. The lower center then carries out the same computation that it would make on the basis of signals from even lower centers when operating in feed-forward mode. In perceptual problems such as the Dalmatian dog, higher centers could suggest edge locations, illusory contours, or cause intensity inputs that represent such hypotheses in response to a combination of the previous analysis of a lower center and top-down information about animals. The lower visual processing area might then be asked to re-compute to see if the augmented image makes logical sense.

For these networks the patterns of activity in cases of 'correct answers' and 'incorrect answers' are so different that they are visually separate classes. The 'aha' feeling that we have when we recall—with much effort—an item from memory and then *know* it is right (or wrong), belongs to same class of phenomenon as visual pop-out.

Although the simulations presented used rate-based units, the use of spiking neurons will add to the capabilities of the networks in at least two ways. First, they should allow memory maintenance (in the presence of synaptic drift or loss) through the mechanism of spike-timing-dependent-plasticity during memory rehearsal. Second, they should permit the parallel search of memories with weak clues through the mechanism of transient synchrony, a synchrony that should be transmitted from the module in which a memory is found to higher modules in the recall pathway. Both of these spiking effects make use of the implicit check-bit idea, but in addition require transient synchrony or near-synchrony in the fundamental computational process. Experimental examination of these ideas through EEG studies or the disruption of synchrony might be feasible.

The connection between the mathematics of the model here and the variables of real neurobiological systems is far from one-to-one. We believe that the model is a dynamical system of the right class, and showed that it displays unexpected and intriguing relationships between phenomena and concepts not generally thought to be related. The focus throughout has been on developing an understanding of the dynamical behavior of the class of neural systems of which the brain is a very complex example. By understanding simple mappings of computational ideas onto the concrete mathematics and simulations of large multi-unit systems, we can examine scientific connections between electrophysiological and psychological or collective phenomena, and go beyond descriptions of brain function that rely chiefly on linguistic metaphor.

Matlab code for the associative memory and the linear programming dynamics of Sudoku are available at http://genomics.princeton.edu/hopfield/   Warning: not well documented

**Acknowledgments**
The author thanks Kevin and Linda Mottram for introducing him to Sudoku puzzles on a windless day while sailing in the Dodecanese, David Eppstein for providing examples of difficult puzzles, and David MacKay for critical comments.  The research was supported in part by National Institutes of Health Grant RO1 DC06104-01.

**References**

Bi, G. and Poo, M. (2001). Synaptic modification of correlated activity: Hebb's postulate revisited. Ann. Rev. Neurosci., **24**, 139-166.

Brody, C. D. and Hopfield, J. J. (2001) What is a moment? Transient synchrony as a collective mechanism for spatiotemporal integration.  Proc. Nat. Acad. Sci. (USA) **98**, 1282-1287.

Crick, F. (1984) Function of the thalamic reticular complex: The searchlight hypothesis, Proc. Nat. Acad. Sci. (USA) **81**, 4586-4950.

Gould, W., (2005) The Times Su Doku Book 1, Harper Collins, London, 76.

Eppstein, D. (2005) Preprint. Nonrepetitive paths and cycles in graphs with application to sudoku. http://arxiv.org/abs/cs.DS/0507053

Hayes, B. (2006) Unwed Numbers: The mathematics of Sudoku, a puzzle that boasts "no math required". Amer. Scientist **94**,12.

Hartwell, L. H., Hopfield, J. J., Leibler, S. and Murray, A. W. (1999) From molecular to modular cell biology, Nature **401** (supp.) C47-C52.

Hopfield, J. J. (1982) Neural networks and physical systems with emergent collective computational properties. Proc. Nat. Acad. Sci. (USA) **79**, 2554-2558.

Hopfield, J. J. (1984) Neurons with graded response have collective computational properties like those of two-state neurons. Proc. Nat. Acad. Sci. (USA) **81**, 3088-3092.

Hopfield, J. J. and Tank, D. W. "Neural" computation of decisions in optimization problems. (1985) Biological Cybernetics **55**, 141-146.

Hopfield, J. J. and Brody, C. D. (2000) What is a moment? "Cortical" sensory integration over a brief interval. Proc. Natl. Acad. Sci. (USA)  **97**, 13919-13924.

James, R. C. (1973) in R. L. Gregory, *The Intelligent Eye*, McGraw-Hill, New York, p. 14.

Jonsson, F. U., Tcheckova, A., Lonner, P. and Olsson, M., J.  (2005) A metamemory perspective on odor naming and identification, Chemical Senses **30**, 353-365.

Lyon, D. C., Jain, N., and Kaas, J. H. (1998)  J Comp. Neurology **401**,109-128.

Metcalfe, J. (2000) Metamemory: Theory and data.  In E. Tulving and F. I. M. Craik (Eds.)  The Oxford Handbook of Memory, pp197-211, New York, Oxford University Press.

Mepham, M. (2004) *Sudoku*, The Daily Telegraph, London.

Nelson, T. O. (1996) Consciousness and metacognition.  American Psychologist **51**, 102-116.

O'Connor, D. H., Wittenberg, G. M., and Wang, S. S.-H. (2005) Graded bidirectional synaptic plasticity is composed of switch-like unitary events. Proc. Nat. Acad. Sci. (USA) **102**, 9679-9684.

Panzeri, S., Rolls, E. T., Battaglia, F. P., and Lavis, R. (2001) Lecture Notes Comp. Sci., **2036**, 320-332.

Peterson, C. C., Malenka, R. C., Nicoll, R. A. and Hopfield, J. J. (1998) All-or-none potentiation at CA3-CA1 synapses.  Proc. Nat. Acad. Sci. (USA) **95**, 4732-4737.

Smith, B.M., Brailsford, S. C., Hubbard, P. M. and Williams, H. P. (1996) Constraints **1,** 119-138.

Tank, D. W. and Hopfield, J. J. (1986) Simple optimization networks: an A/D converter and a linear programming circuit. IEEE Circuits and Systems **CAS-33,** 533-541.

Treisman, A. M. and Gelade, G. (1980) A feature integration theory of attention. Cog. Psych. **12**, 97-136.

Treisman, A. M. (1986). Features and objects in visual processing. Scientific American, **225**, 114B-125.

D. C. van Essen, C. H. Anderson, and D. J. Fellerman (1992) Information processing in the primate visual system: an integrated systems perspective. Science **255**, 419-423.

van Essen, D. C. (1985) Functional organization of primate visual cortex. . In A. Peters and E.G. Jones, eds., Cerebral Cortex, vol. 3, Plenum Press, New York, pp 259--329.

van Essen, D. C. (2004) Organization of visual areas in macaque and human cerebral cortex.  In *The Visual Neurosciences*, Vol. 1, L. M  Chalupa and J. S. Werner, eds., MIT Press, 507-521.

Yato, T., and Seta, T. (2002). Complexity and completeness of finding another solution and its application to puzzles. http://www-imai.is.s.u-tokyo.ac.jp/~yato/data2/SIGAL87-2.pdf